\documentclass[aps,prb,twocolumn,amsmath,amssymb,superscriptaddress,longbibliography]{revtex4-2}
\usepackage[dvips]{graphicx}
\usepackage{graphicx}
\usepackage{soul,xcolor}
\setstcolor{red}

\usepackage[utf8]{inputenc}
\usepackage{dcolumn}
\usepackage{physics}
\usepackage{bm}
\usepackage{graphics}
\usepackage{epsfig,color}
\usepackage[breaklinks,colorlinks,bookmarks=false,citecolor=red,linkcolor=blue,urlcolor=blue]{hyperref}
\graphicspath{{Fig/}} 

\begin{document}
\title{Quantum correlations in the frustrated XY model on the honeycomb lattice}

\author{Sahar Satoori}
\affiliation{Department of Physics, University of Guilan, 45196-313, Rasht, Iran.}

\author{Saeed Mahdavifar}
\email[]{smahdavifar@gmail.com}
\affiliation{Department of Physics, University of Guilan, 45196-313, Rasht, Iran.}

\author{Javad Vahedi}
\email[]{j.vahedi@jacobs-university.de}
 \affiliation{Jacobs University, School of Engineering and Science, Campus Ring 1, 28759 Bremen, Germany.}

\date{\today}
\begin{abstract}
We consider the spin-1/2 XY frustrated antiferromagnetic Heisenberg honeycomb model. There is an unclear intermediate region in the ground state phase diagram of the model.  The most recognized phases are the quantum spin-liquid (QSL) and the antiferromagnetic Ising ordering. From the viewpoint of the quantum correlations, the QSL phase is expected to be entangled. Motivated by this fact, we have calculated the concurrence, the quantum discord (QD), and the entanglement entropy, by using numerical Lanczos and density matrix renormalization group (DMRG) methods.  Our results explicitly show that the intermediate region should be entangled supporting the QSL phase.
\end{abstract}

\maketitle
\section{INTRODUCTION}\label{sec1}
The exploration of the quantum phase transitions~\cite{a1,a6} in low-dimensional magnets is one of the essential topics in condensed matter physics. Quantum phase transition arises from quantum fluctuations at absolute zero temperature, where their effects vary depending on the spin quantum number, coordination number, and the presence of frustration.
\par
Frustrated low-dimensional spin systems play a key role in understanding the quantum phases of a matter~\cite{sa1}. Frustration in magnets arises from the incompatibility of interactions between spins that cannot be simultaneously satisfied, giving rise to the macroscopic degeneracy of the ground state. In recent decades both theoretical and experimental, the realization of exotic quantum phases in frustrated one- and two-dimensional (2D) systems has been one of the appealing topics. As an example, spin-$1/2$ antiferromagnetic frustrated isotropic Heisenberg chains have been studied extensively, and it is well known to display a quantum phase transition from the Luttinger liquid phase to a dimer phase at a quantum critical frustration value\cite{sa2,sa3,sa4}. 
\par
In two-dimensional frustrated models, most studies have been focusing on triangular, square, kagome, and honeycomb lattices\cite{sa5}. The main reason for studying frustrated 2D models is the existence of an exotic quantum phase called quantum spin-liquid (QSL)\cite{QSL1, QSL2, QSL3}. The QSL is a superposition of valence-bound states and a prototypical example of a ground state with many-body entanglement. The discovery of the high-temperature copper oxide superconductors~\cite{sa7, sa8} and Anderson's report of the existence of the RVB grand state in antiferromagnetic Heisenberg model on a hexagonal lattice\cite{sa9} equipped the field of the research to QSL phase.  
\par
Among the frustrated systems on the two-dimensional lattices, the spin-$1/2$ honeycomb lattice has received a lot of attention. First of all, in the absence of frustration, the isotropic Heisenberg honeycomb model exhibits the long-range N\'eel order\cite{H1,H2,H4}. In 2001,  using the numerical Lanczos method, a quantum phase transition into the QSL or dimer phase has recognized in the presence of the frustration\cite{FH1} and confirmed by different numerical and analytical methods ~\cite{FH2,FH3,FH4,FH4-1,FH4-2,FH4-3,FH4-4,FH4-5} and also observed experimentally\cite{FH5,FH6}.
\par
In addition to the isotropic Heisenberg model, the ground-state phase diagram of the spin-$1/2$ frustrated honeycomb XY model is also investigated extensively. First, Varney et. al. studied the model via the numerical exact diagonalization method\cite{FXY1}. Concentrating on the fidelity parameter in finite-size latices, they recognized four distinct phases separated by three critical quantum phase transitions in clusters with $N=24$ spin-$1/2$ particles. In addition to the N\'eel  and a spin-wave with $120^{\circ}$ order, two intermediate phases: (1) a QSL and (2) an exotic spin wave state are suggested. The stability of the QSL phase against various perturbations is shown by the numerical Lanczos method\cite{FXY2}. Applying variational Monte Carlo, it is found that in the intermediate region, the ordered phases lose energy to an exotic fractionalized partonic wave function that is consistent with the suggested gapped QSL phase\cite{FXY3}. In addition, using the extended path integral Monte Carlo simulations, it is shown that the QSL is stable in the intermediate region instead of the Ising phase suggested by the density matrix renormalization group\cite{FXY8}. On the other hand, in a different analytical work, using a variational approach based on the Jastrow wave functions, the existence of the QSL phase in the intermediate region has not been confirmed\cite{FXY4}. 
\par
However, numerical results based on the density matrix renormalization group(DMRG) method put forward more challenges and puzzles in the field.  Instead of the QSL phase,  the antiferromagnetic Ising phase with staggered magnetization polarized along the $z$-direction is found in the intermediate frustration regime\cite{FXY5, FXY6}. They obtained a ground state energy much lower than the proposed QSL and vanishing topological entanglement entropy. Later,  using the series expansion method, it is shown that in the vicinity of the first critical point, the nearest-neighbor $zz$-correlations rise to become equal to the nearest-neighbor $xy$-correlations, which provides indirect support for existing of the Ising phase in the intermediate region\cite{FXY7}.  Recently, and based on DMRG, the effect of a three-spin chiral term is studied and it is found that the Ising phase in the intermediate region remains stable in presence of such a perturbation\cite{FXY11}. The boson-vortex duality has also been used to examine the mentioned model\cite{FXY9}. It has shown that by condensing one of the two vortex flavors, Ising ordered phase is seen in the intermediate region while condensing both vortex flavors, the QSL will be replaced. Additionally, using bosonic dynamical mean-field theory, an emergent chiral spin state in the intermediate frustration regime is recognized instead of Ising and QSL phases\cite{FXY10}. A study based on the Chern-Simons fermionization of spin-$1/2$ operators is shown a composite structure of ground-state order in the intermediate region, characterized by the coexistence of out-of-plane long-range N\'eel and chiral spin-liquid ordering in-plane\cite{FXY12}.

\begin{figure}[t]
\includegraphics[width=0.5\textwidth]{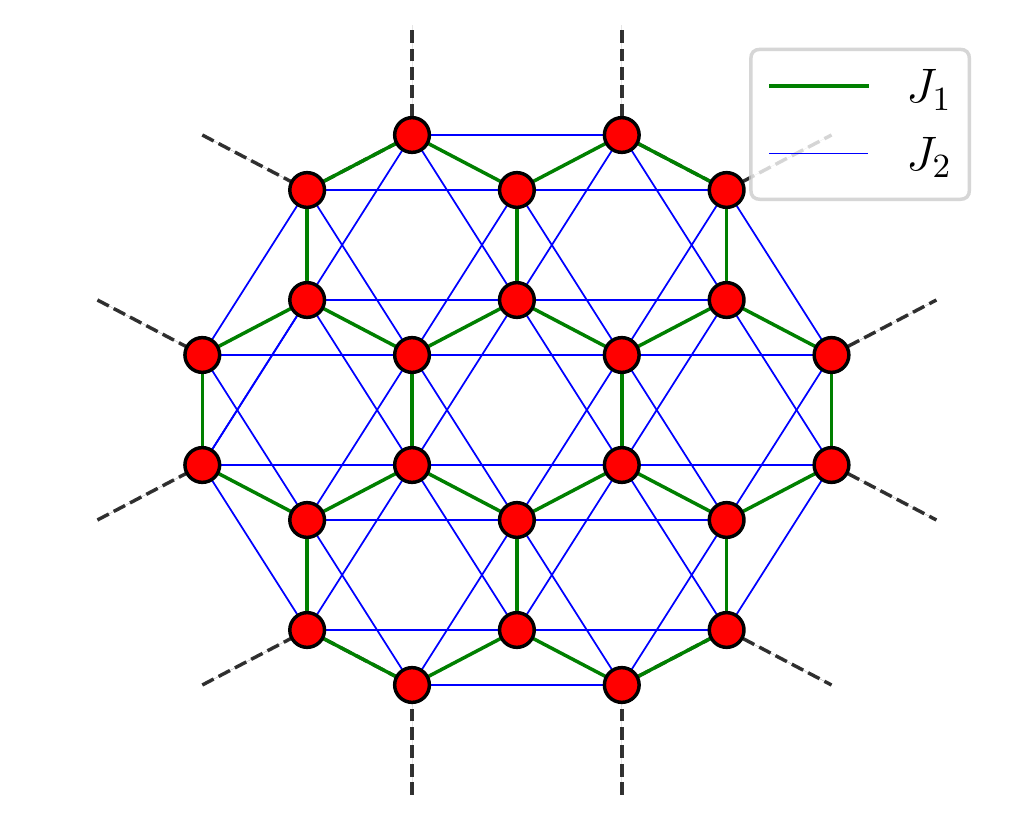}
\caption{The cluster of $N=24$ spins with hexagonal  symmetry. Green and blue links show coupling between nearest and next-nearest neighbor spins, respectively. A twist periodic boundary condition is implement.}
\label{fig0}
\end{figure}
What we conclude is that the ground-state phase diagram of the 2D spin-$1/2$ XY frustrated antiferromagnetic Heisenberg honeycomb model is ambiguous and controversial. Then motivated by the role of quantum correlations in identifying the critical quantum phase transition points, both associated with the symmetries and topological orders, we provide some numerical analyses on the ground-state phase diagram of the  spin-$1/2$ XY frustrated honeycomb model. The study of quantum phase transitions through quantum entanglement concepts provides a new way to understand strongly correlated systems near criticality\cite{Satoori_2022}. In this work, using the numerical Lanczos and DMRG methods, we calculate the concurrence, the quantum discord (QD), and the entanglement entropy on finite clusters (please see Fig.~\ref{fig0}). Results show that the frustration cannot induce the entanglement between the next-nearest neighbor (NNN) pairs of spins. Nearest neighbor (NN) pairs are entangled in the intermediate region. QD exists between the NN and NNN pair of spins in the mentioned intermediate region. By analyzing the first derivative of the concurrence and the QD, all three critical points are recognized in complete agreement with the fidelity results\cite{FXY1}. By dividing the cluster into two subsystems and using the entanglement entropy concept, we show that the system is in a many-body entangled state in the intermediate region, which supports the existence of the QSL phase.  
\par
The paper is organized as follows. In Sec. II, we introduce the spin-$1/2$ frustrated antiferromagnetic $XY$ model on the honeycomb lattice with the nearest and next-nearest neighbors exchange interaction. In Sec. III, We briefly review the concepts of concurrence, quantum discord (QD), and entanglement entropy from the information theory point of view.  In Sec. IV, we represent our results on the behavior of the concurrence, QD, and entanglement entropy between the NNs and NNNs pair of spins in terms of the frustrated parameter $\alpha$. A summary is given in  Sec. V.

\section{Model} \label{sec2}
We consider the  2D spin-$1/2$ XY frustrated antiferromagnetic Heisenberg honeycomb model. The Hamiltonian of the model is defined as
\begin{equation}
    H=J_1 \sum_{<i,j>} (S_i^x  S_j^x+ S_i^yS_j^y)+J_2\sum_{\ll i,j\gg} (S_i^x  S_j^x+ S_i^y S_j^y)
\end{equation}

where $S_i^{x(y)}$ is the in-plane component of the spin-$1/2$ operator on the $i$-th site and $J_1$ and $J_2$ are the antiferromagnetic coupling constants between NN and NNN spins. $<i,j>$ and $\ll i,j\gg$  indexes sum over all of the NN and NNN spins, respectively. $\alpha=\frac{J_{2}}{J_{1}}$ is defined as the frustration  parameter.  As mentioned, the ground-state phase diagram of the model has been investigated within various techniques at the quantum level. In 2011, Varney and co-workers investigated the spin-$1/2$ frustrated antiferromagnetic $XY$ model on the honeycomb lattice using the exact diagonalization method on finite clusters~\cite{FXY1}. System at zero temperature exhibits four different phases as: (I) in-plane Neel phase in the region $\alpha<\alpha_c\sim 0.21$,  (II) the gapless QSL phase in the region $\alpha_{c_1}\sim0.21<\alpha<\alpha_{c_2}\sim 0.35$, (III)  the collinear spin wave phase in the region $\alpha_{c_2}\sim0.35<\alpha<\alpha_{c_3}\sim 1.32$, and (IV) $120^{\circ}$ order  in the region $\alpha>\alpha_{c_3} \sim 1.32$. As we have pointed out in the previous part, the nature of the system in the intermediate region $\alpha_{c_1}<\alpha<\alpha_{c_2}$ is unclear, and studies on it are ongoing.
\par
In the following, by focusing on observables borrowed from the quantum information community, namely the concurrence, the QD, and the entanglement of entropy, we try to explore the existence of the four phases and the precise location of all the critical points and also comment on the nature of the intermediate doubt region from the many-body entanglement point of view. 

\section{Quantum Correlations} \label{sec4}

The fields of statistical mechanics, condensed matter physics, and quantum information theory share a common interest in the study of interacting quantum many-body systems. Entanglement is one of the interesting subjects in physics and plays an important role in all these areas ~\cite{E1,E2,E3,E4}. In a bipartite system, entanglement occurs when we cannot obtain the state of each subsystem independently from the state of the composite system. In other words, the entanglement explains how many quantum effects we can observe and use to control one subsystem by another. Exotic ground state quantum states such as QSL ~\cite{6,7}, topological phases  ~\cite{8,9,9b}, and many-body localized systems ~\cite{10} find their hallmarks in their quantum correlation features. In recent years, it is known that entanglement in quantum many-body systems can be accessible in experiments  ~\cite{13,14, 15, 16}. There are different methods for measuring quantum correlations. Among them, we can mention quantities such as the concurrence, the QD, and the entanglement entropy, which can be applied to detecting many-body entanglement.

\subsection{Concurrence} \label{subsecA}
Here, we present a brief review of a tool called concurrence  for measuring the entanglement. The concurrence is an entanglement monotone defined for a mixed state of two spin-$1/2$ particles  as\cite{E4-1}:
\begin{equation}
C_{ij}=C(\rho_{ij})=Max \{  0, \lambda_{0}- \lambda_{1}- \lambda_{2}- \lambda_{3} \},
 \end{equation}
where $ \lambda_{i}$'s are the eigenvalues of the reduced density matrix $R_{ij}=\sqrt{\rho_{ij}  \tilde{\rho}_{ij}} $ in which $ \tilde{\rho}$ is spin-flipped state $\rho$ and is written as
\begin{equation}
 \tilde{\rho}_{ij}=(\sigma^{y}_{i} \otimes \sigma^{y}_{j})~ \rho^{*\ }_{ij} (\sigma^{y}_{i} \otimes \sigma^{y}_{j}). 
 \end{equation}
For a pair of spin-$1/2$ particles located at sites $i$ and $j$, the concurrence can be determined by the corresponding reduced density matrix $\rho_{ij}$, 

\begin{eqnarray}
\rho_{ij}=\left(
\begin{array}{cccc}
X_{ij}^{+} & 0 & 0 & F_{ij}^{*} \\
0 & Y_{ij}^{+} & Z_{ij}^{*} & 0 \\
0 & Z_{ij} & Y_{ij}^{-} & 0 \\
F_{ij} & 0 & 0 & X_{ij}^{-} \\
\end{array}
\right),
\label{density matrix2}
\end{eqnarray}
where 
\begin{eqnarray} 
X_{ij}^{+}&=& \langle(1/2+S_{i}^{z})(1/2+S_{j}^{z})\rangle,\nonumber\\
Y_{ij}^{+}&=& \langle(1/2+S_{i}^{z})(1/2-S_{j}^{z})\rangle,\nonumber\\
Y_{ij}^{-}&=& \langle(1/2-S_{i}^{z})(1/2+S_{j}^{z})\rangle,\\
X_{ij}^{-}&=& \langle(1/2-S_{i}^{z})(1/2-S_{j}^{z})\rangle,\nonumber\\
Z_{ij}&=& \langle S_i^{+}S_{j}^{-}\rangle,\nonumber\\
F_{ij}&=& \langle S_i^{+}S_{j}^{+}\rangle
\label{dm2}
\end{eqnarray}
and $\langle...\rangle$ represents the expectation value on the ground state of a quantum system. Finally, the concurrence is given by the following expression,
\begin{eqnarray}
C_{i,j} &=& \max{\{0,2 (C_1,C_2)\}}, \nonumber \\
C_1&=&|Z_{ij}|-\sqrt{X_{ij}^{+}X_{ij}^{-}}, \nonumber \\
C_2&=&|F_{ij}|-\sqrt{Y_{ij}^{+}Y_{ij}^{-}}~.
\label{C1}
\end{eqnarray}

\subsection{Quantum Discord} \label{subsecB}
In information theory, mutual information is defined as a measure of the mutual dependence between the two random variables. On the other hand, mutual information describes how much information can be determined about a random variable, by knowing the value of a correlated other random variable. To capture the quantum correlations presented in a bipartite state that are not explored by the concurrence, one can also calculate the QD ~\cite{52,53}. QD is the difference between quantum and classical correlations. To understand QD, we consider two spins located at sites $i$ and $j$. In classical information theory, the joint entropy, the amount of information obtained from observation of $S_i$ and $S_j$ at the same time, is defined as
\begin{equation}
H(i , j)=- \sum p(i , j) log_2 p(a , j),
 \end{equation}
where $ p(i,j)$ is the joint probability distribution, which characterized the total correlation between two spins $S_i$ and $S_j$. In addition, the conditional entropy is given by
\begin{equation}
H(i | j)= H(i , j) - H(j),
 \end{equation}
now, the mutual information is expressed as
\begin{equation}
{\cal I}(i : j)= H(i) + H(j) - H(i , j),
 \end{equation}
also, it can be written as 
\begin{equation}
{\cal I}(i : j)= H(i)  - H(i | j).
 \end{equation}
In classical information theory, these two phrases are equivalent. In quantum information theory, the Shannon entropy and probability distribution are replaced by the Von-Neumann entropy and density matrix, respectively. Thus, the quantum mutual information for a bipartite quantum state $\rho_{ij}$ can be redefined as
\begin{equation}
{\cal I}(\rho_{ij})= S(\rho_{i}) + S(\rho_{j}) + \sum\limits_{\alpha  = 0}^3 {{\lambda _\alpha }} \log ({\lambda _\alpha }),
 \end{equation}
and
\begin{equation}
{\cal C}(\rho_{ij})= S(\rho_{i})  -  S(\rho_{i} | \rho_{j}),
 \end{equation}
is classical correlation. $S(\rho_{i} | \rho_{j})$ is a quantum generalization of the conditional entropy and should be measured over all possible states of the subsystem $S_j$.
Unlike classical information, ${\cal I}(\rho_{ij})$ and ${\cal C}(\rho_{ij})$ are not the same, and the difference between them is so-called QD, 
\begin{equation}
QD={\cal I}(\rho_{ij})  - {\cal C}(\rho_{ij})
 \end{equation}
The parameters needed to calculate the total and classical correlations are provided from the elements of the reduced density matrix. Details of calculations of the QD are presented in Appendix\ref{apxa}. 

\begin{figure}[t]
\centerline{\psfig{file=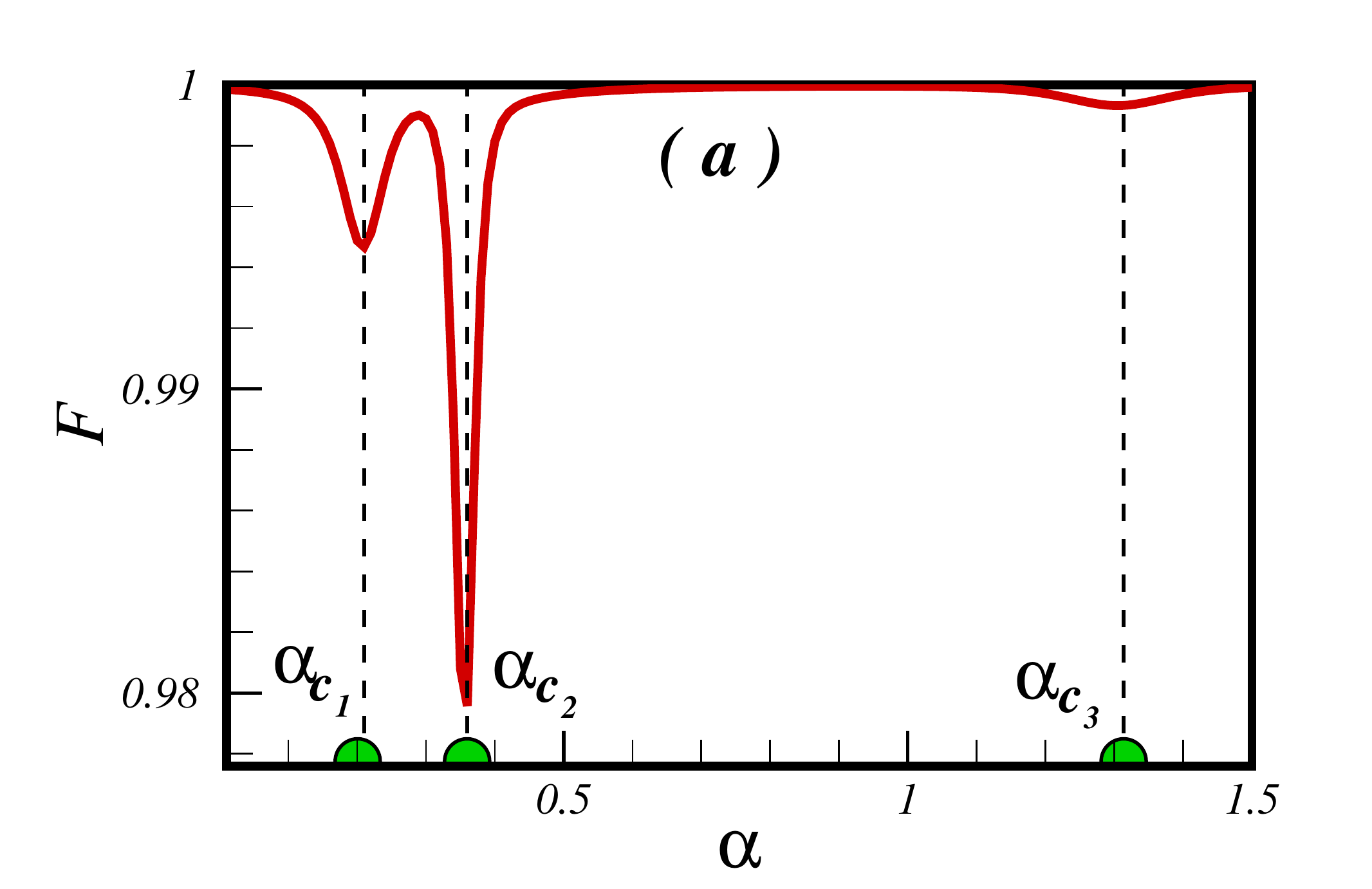,width=1.9in}\psfig{file=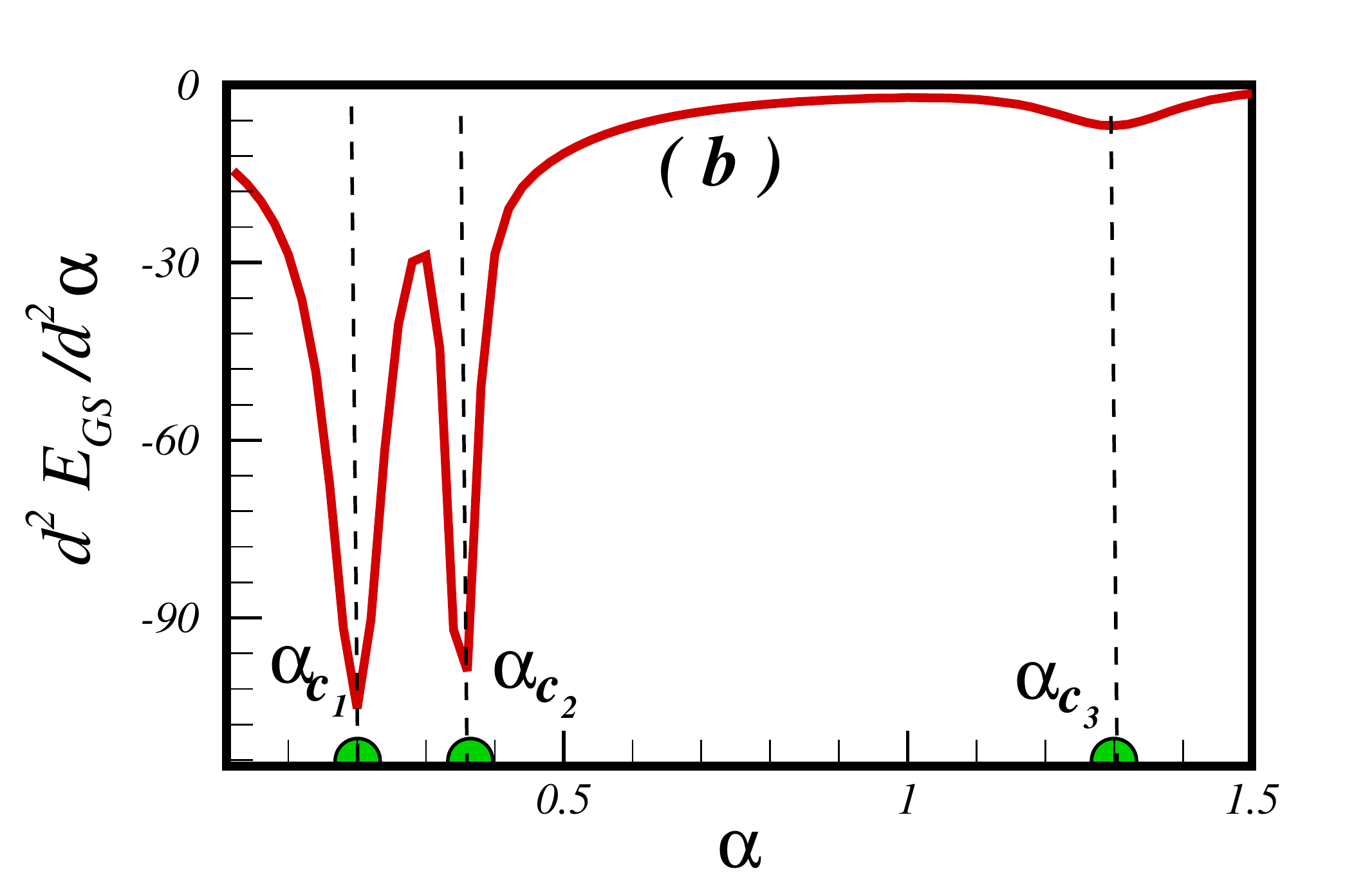,width=1.9in}}
\caption{ Signature of the quantum critical points is clearly seen in:  (a) the fidelity metric as a function of frustration parameter $\alpha$ and (b) the second derivative of the ground state energy with respect to the frustration parameter for cluster with $N=24$ spins.   }
\label{fig1}
\end{figure}

\subsection{Entanglement Entropy} \label{subsecC}

The entanglement entropy (EE) define as the von Neumann entropy of a reduced density matrix of a subsystem~\cite{EE1, EE2, EE3, EE4, EE5}. It  measures the degree of correlation of the ground state between two subsystems $A$ and $B$ in a composite system. The EE is known as a novel tool, very useful in characterizing quantum phases with many-body entanglement and phase transitions in a broad set of lattice models relevant for condensed matter systems. The EE  given by
\begin{equation}
S_{A}=- Tr[ \rho_{A} log (\rho_{A})]
 \end{equation}
where
\begin{equation}
\rho_{A}= Tr_{B} (\rho_{AB})
 \end{equation}
 is the reduced density matrix of subsystem $A$ taken by tracing over the rest of the system, $B$. The EE is suggested to obey the so-called area law, i.e. in a $d$-dimension models  it scales with the surface of the subsystem $A$ as
 
\begin{equation}
S_{A}= \alpha~ l^{d-1}+...,
 \end{equation}
where the ellipsis represents terms that will be vanished in the limit $l \longrightarrow \infty$. Note that the EE only in non-critical systems  obeys a strict area law, and critical systems weakly violate the area law by a multiplicative logarithmic correction.


\section{Numerical results} \label{sec3}
In this section, we apply the two most numerical approaches, the numerical Lanczos~\cite{Lanczos_1950} and DMRG~\cite{dmrg1992} methods, for computing the ground-state eigenvector of the system and extracting quantum correlations on honeycomb finite-size lattices. We should mention that the DMRG approach was originally designed for one dimension, but going to the 2D geometry makes it difficult as the computational effort scales exponentially with the width of the system\cite{Stoudenmire2012}. However, the lack of alternative approaches (as the quantum Monte Carlo has shown limitations for systems with the sign problem \cite{Troyer2005} or exact diagonalization methods) makes DMRG a powerful method for studying complex 2D systems\cite{Misguich2021,Scholl2021,Samajdar2021}. The DMRG calculations in this paper were performed using the ITensor C++ library (version 3.1)~\cite{itensor}. Periodic boundary conditions are considered for honeycomb lattices with finite clusters. Throughout the paper, we mainly consider symmetric hexagonal-shaped clusters (see the sketch in  Fig.~\ref{fig0}) as we noticed a less finite-size effect. Other clusters with different shapes were also explored, see Fig.\ref{App1}, and the results are presented in Appendix~\ref{apxb},  and Fig.~\ref{App2}.  The ground state of the system is also calculated by the Lanczos algorithm (or DMRG method) and subsequently, quantum correlations as the concurrence, the QD, and the EE are obtained. 
\par.pdf
As the main focus of this work is Ref.\cite{FXY1},  it is worth recovering their results using the  quantum ground-state fidelity, $F= \langle \psi_{Gs} (\alpha)| \psi_{Gs} (\alpha+\delta \alpha)| \rangle$,  and the second derivative of the ground-state energy with respect to the frustration parameter.   Fig.~\ref{fig1} shows results based on the Lanczos and DMRG techniques for a cluster with $N=24$ spins. It can be seen, the fidelity remains almost one except in the vicinity of the quantum critical points where it falls. The same behavior is observed in the plot of the second derivative of the ground-state energy, which beliefs to capture the second-order quantum phase transition types. The critical points are obtained as: $\alpha_{c_1}=0.214\pm 0.002$,  $\alpha_{c_2}=0.352\pm 0.002$,  $\alpha_{c_3}=1.272\pm 0.02$, in complete agreement with numerical and analytical approaches. 
\begin{figure}[t]
\centerline{\psfig{file=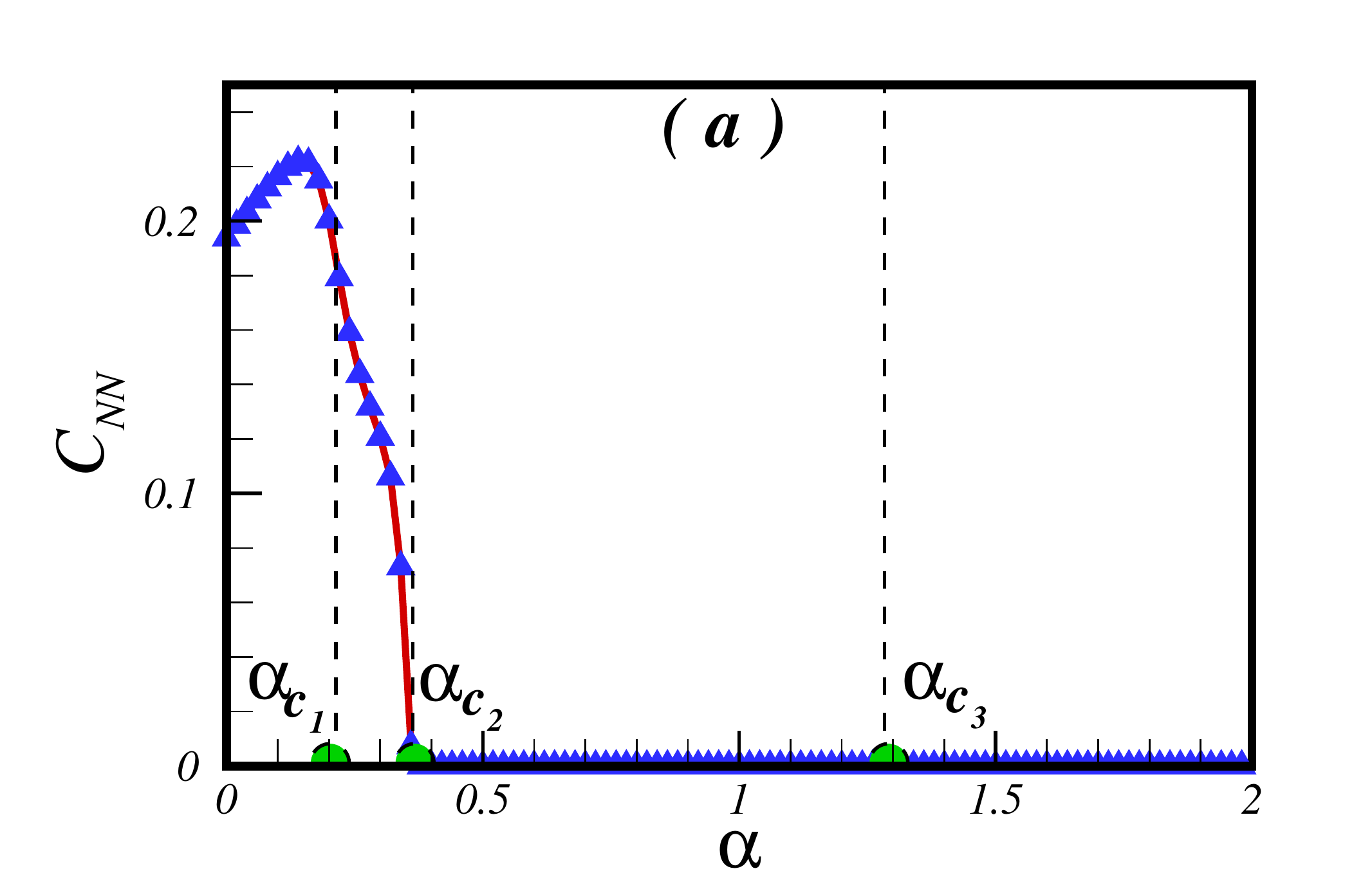,width=1.9in}\psfig{file=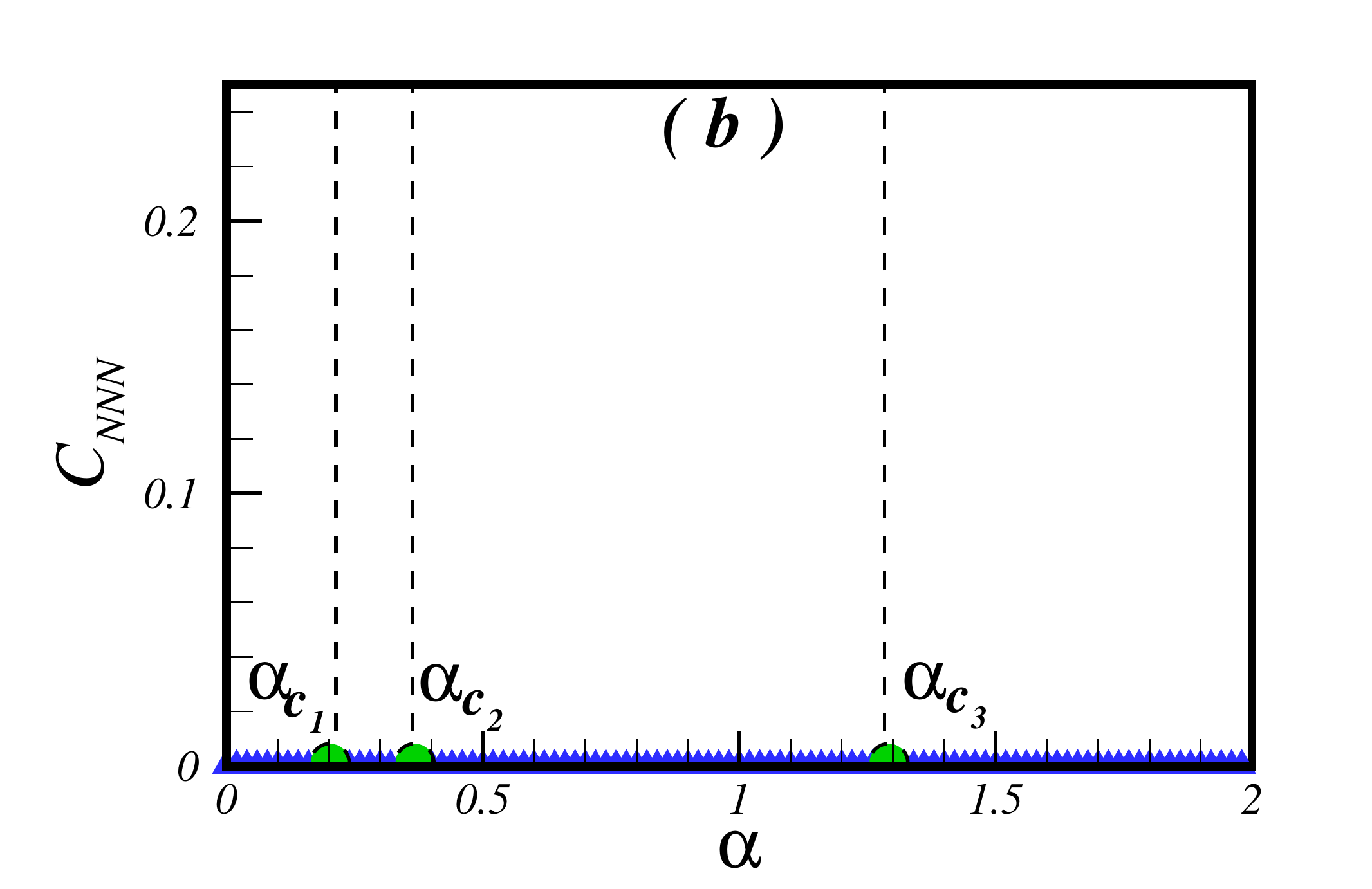,width=1.9in}}
\centerline{\psfig{file=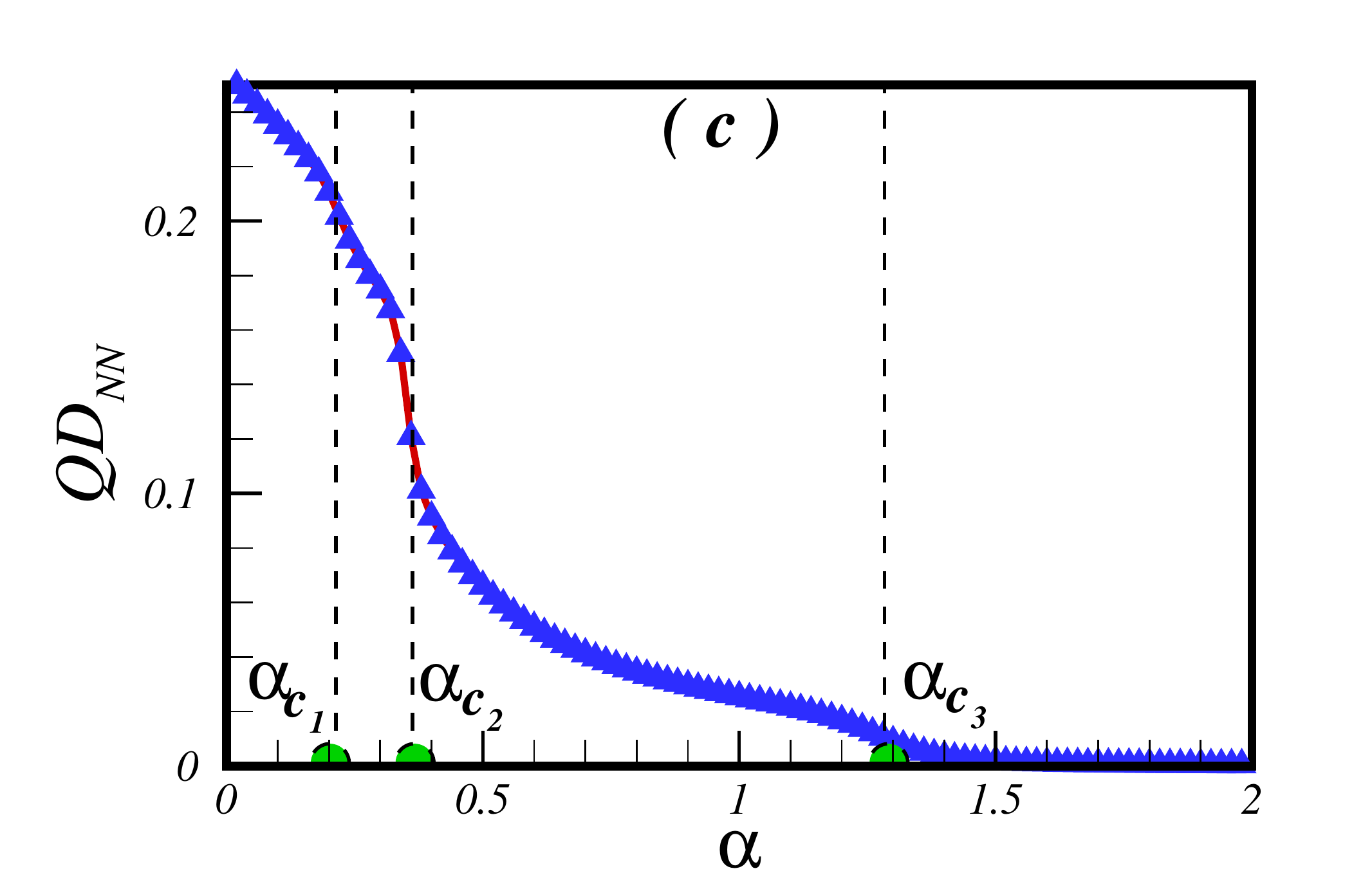,width=1.9in}\psfig{file=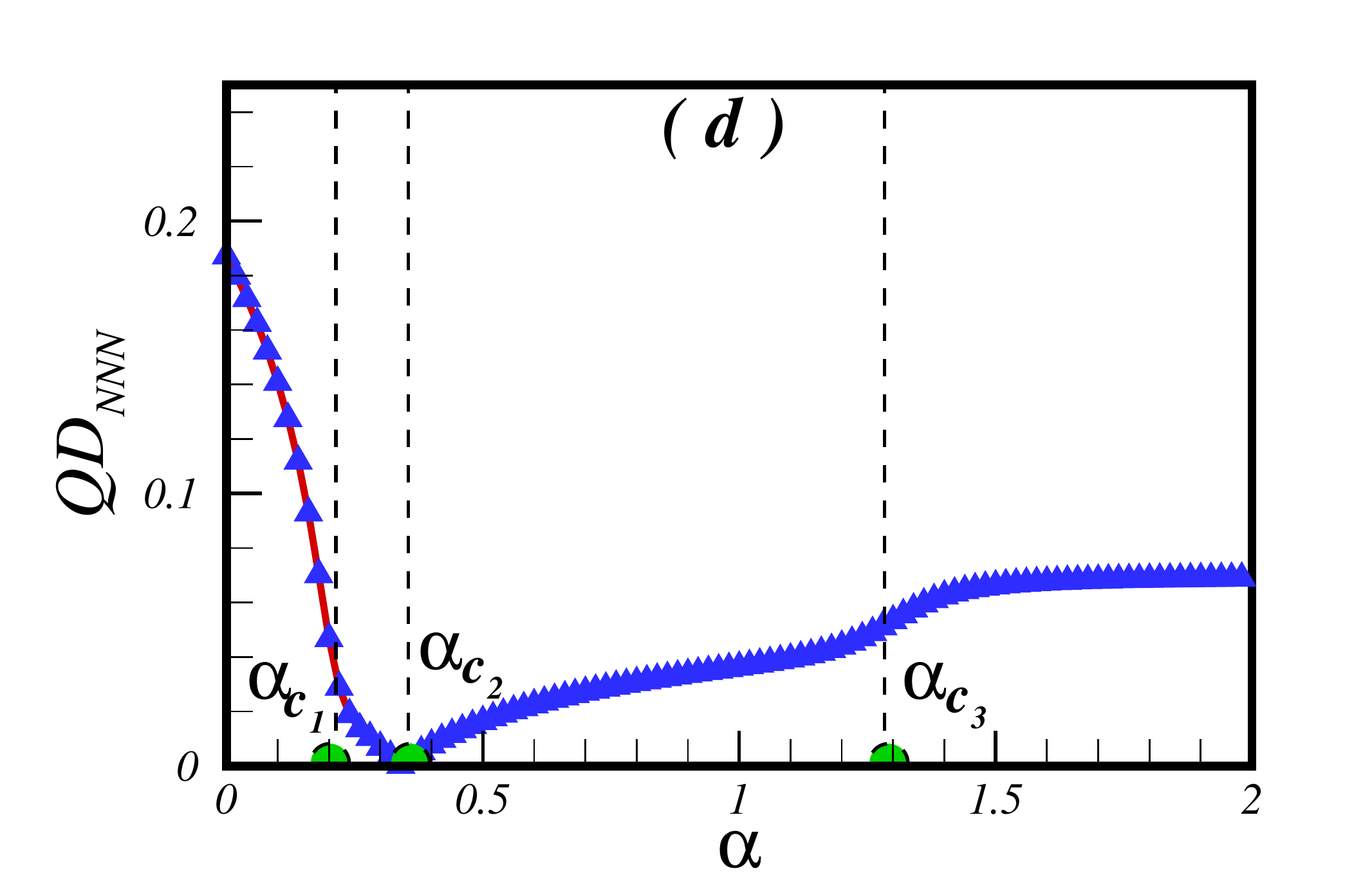,width=1.9in}}
\caption{ The concurrence and the QD as a function of  the frustrated parameter $\alpha$  between the NN ((a) and (c)) and the NNN  ((b) and (d)) pair of spins in a cluster of $N=24$ spins.}
\label{fig2}
\end{figure}

\par
Fig.~\ref{fig2} depicts the behavior of the concurrence and the QD between a pair of spins as a function of the frustration parameter $\alpha$ for a cluster of $N=24$. As shown in Fig.~\ref{fig2}-(a), the NN pair of spins are entangled in the absence of frustration. As soon as the interaction between the NNN pair of spins is applied, $C_{NN}$ increases up to a value close to the first critical point, which is the boundary point of the QSL phase. This shows that weak frustration increases quantum fluctuations and thus entanglement between NN pair spins enhances through mixing mechanisms. By increasing the frustration more, $C_{NN}$ decreases in the critical region, and by entering the QSL phase, the mentioned concurrence decreases monotonically and vanishes at the second critical point $\alpha_{c_2}$, which is the boundary point of the collinear spin-wave phase. In the collinear and $120^{\circ}$ ordered phases, the NN pair of spins are not entangled.  A particle in absence of the frustration is in interaction with three particles. But, frustration increases the number of interactions of a particle with others to nine. This causes less quantum fluctuations and therefore decreasing and vanishing entanglement in the mentioned model. In addition to the NN pair of spins, NNN spins are not entangled and the frustration parameter does no effect on the entanglement between the NNN pair of spins, as is seen in Fig.~\ref{fig2}-(b).
\par
The QD is also calculated for the same pair of spins and the results are presented in Fig.~\ref{fig2} (c)-(d). As it is apparent, the QD exists between both NN and NNN spins in the absence of the frustration. Increasing frustration parameter from zero and in the N\'eel phase, the QD between NN and NNN pair of spins start decreasing, but with different slopes. QD between NNN spins decreases faster than the other one. With passing the first critical point $\alpha_{c_1}$, we have not seen any clear signature on the QD between the NN and NNN pair of spins. However, ${\rm QD}_{\rm NN}$ drops at the second critical point $\alpha_{c_2}$. We should  note that ${\rm QD}_{\rm NNN}$ becomes almost negligible at the mentioned critical point. In the collinear spin-wave phase, ${\rm QD}_{\rm NN}$ and ${\rm QD}_{\rm NNN}$ behave inversely. Although QD between NN spins decreases by enhancing the frustration, ${\rm QD}_{\rm NNN}$ shows increasing behavior. In other words, as the NN pair of spins lose the quantum correlations detectable by QD, the NNN pair of spins gets the mentioned quantum correlations. This can be explained as follows, the spin-wave structure of the magnetic phase to stabilize itself needs to compensate correlations between spins over the nearest ones. During this process for the model of study, this provides by the second neighbors, and quantum correlations in form of QD evolve between the NN and NNN spins inversely by raising the frustration parameter $\alpha$.  Finally,  at the third quantum critical point  $\alpha_{c_3}$, QD between NN becomes almost zero and shows asymptotic behavior in the $120^{\circ}$ order phase. On the other hand, ${\rm QD}_{\rm NNN}$ saturates to a finite value and remains almost constant in the $120^{\circ}$ ordered phase. We give a short explanation based on the reduced density matrix formalism  for a toy three spins model in Appendix~\ref{apxc}. 
\par
\begin{figure}[t]
\centerline{\psfig{file=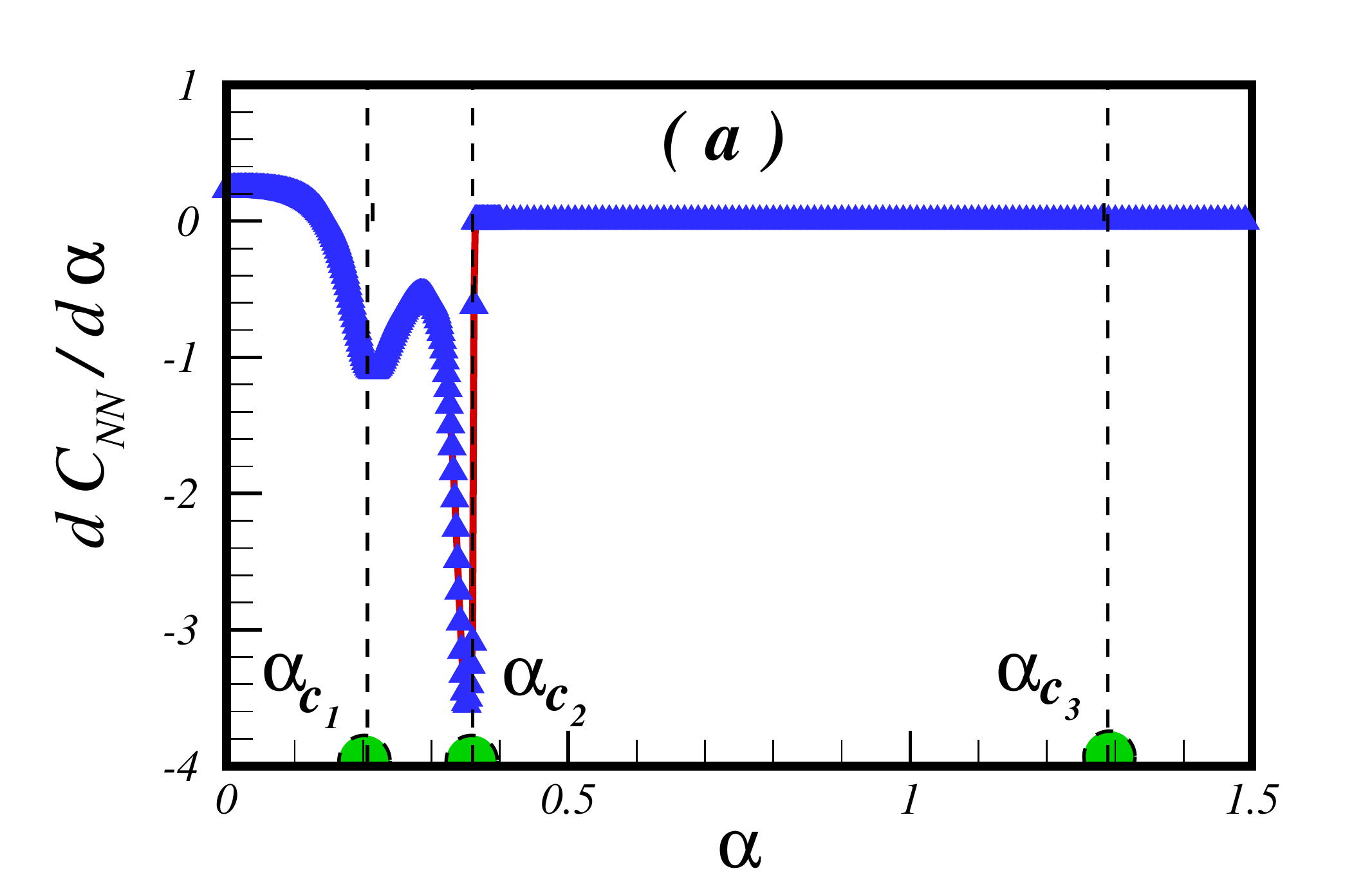,width=1.9in}\psfig{file=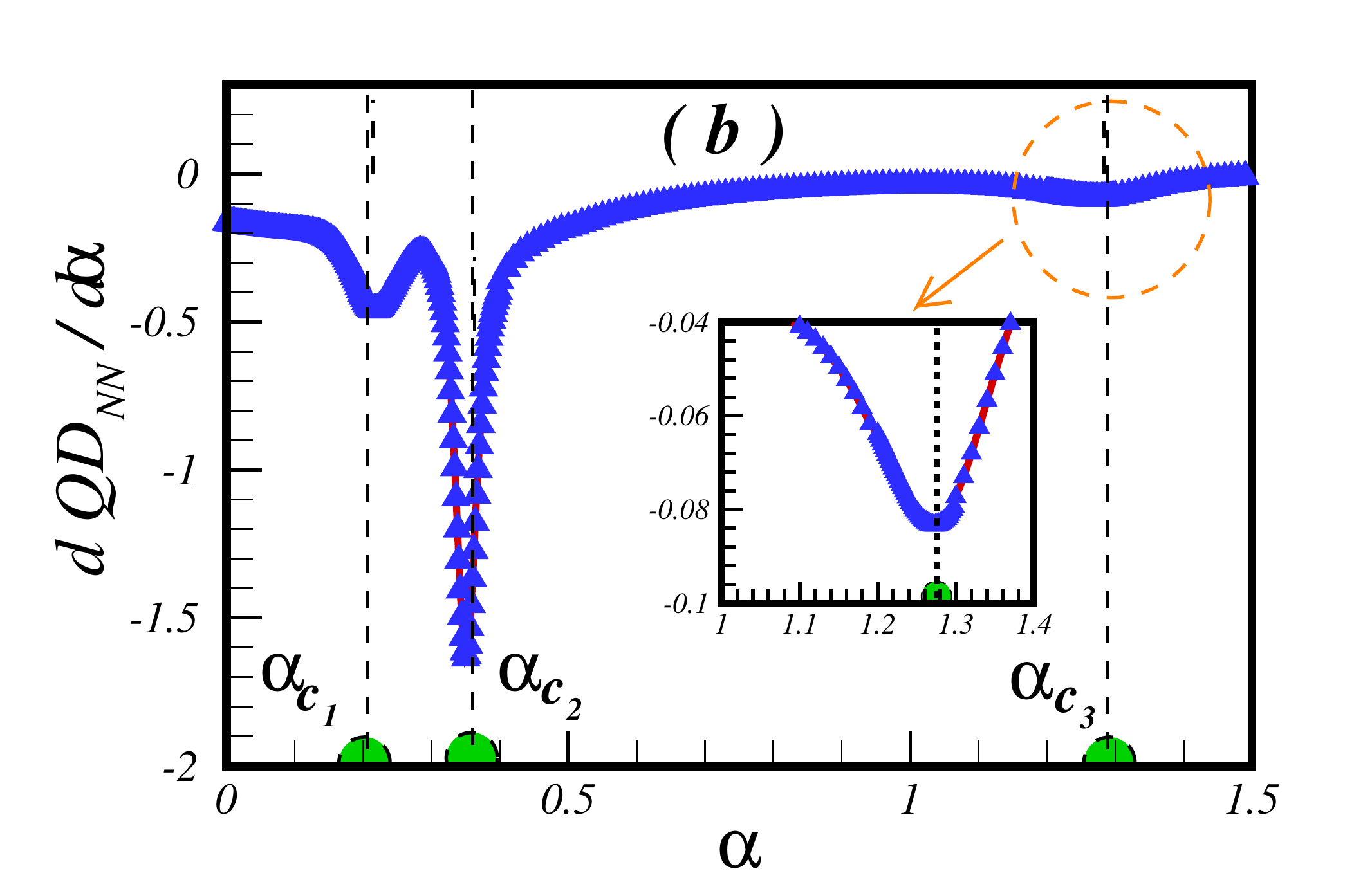,width=1.9in}}
\caption{The first derivative of (a) the concurrence and (b) the QD between NN pair of spins with respect to the frustration parameter $\alpha$.}
\label{fig4}
\end{figure}

Although the quantum phase transition describes an abrupt change in the ground state of a many-body system and can be detected from the state functions of the many-body system, then the derivation of observables is also expected to catch good information about the quantum phase transitions. Here, to probe the quantum phase transition points, we have numerically computed the first derivative of the concurrence and the QD between the NN pair of spins, and the results are shown in Fig.~\ref{fig4}. It seems that the sharp changes (or drops) of the first derivative of the concurrence and the QD between the NN pair of spins ($\alpha_{c_1}=0.214\pm 0.002$,  $\alpha_{c_2}=0.352\pm 0.002$,  $\alpha_{c_3}=1.272\pm 0.02$) can be ascribed to a dramatic change in the structure of the ground-state during the quantum phase transitions. Away from these critical points, the first derivative of the concurrence and the QD almost remain constant.

\begin{figure}[t]
\centerline{\psfig{file=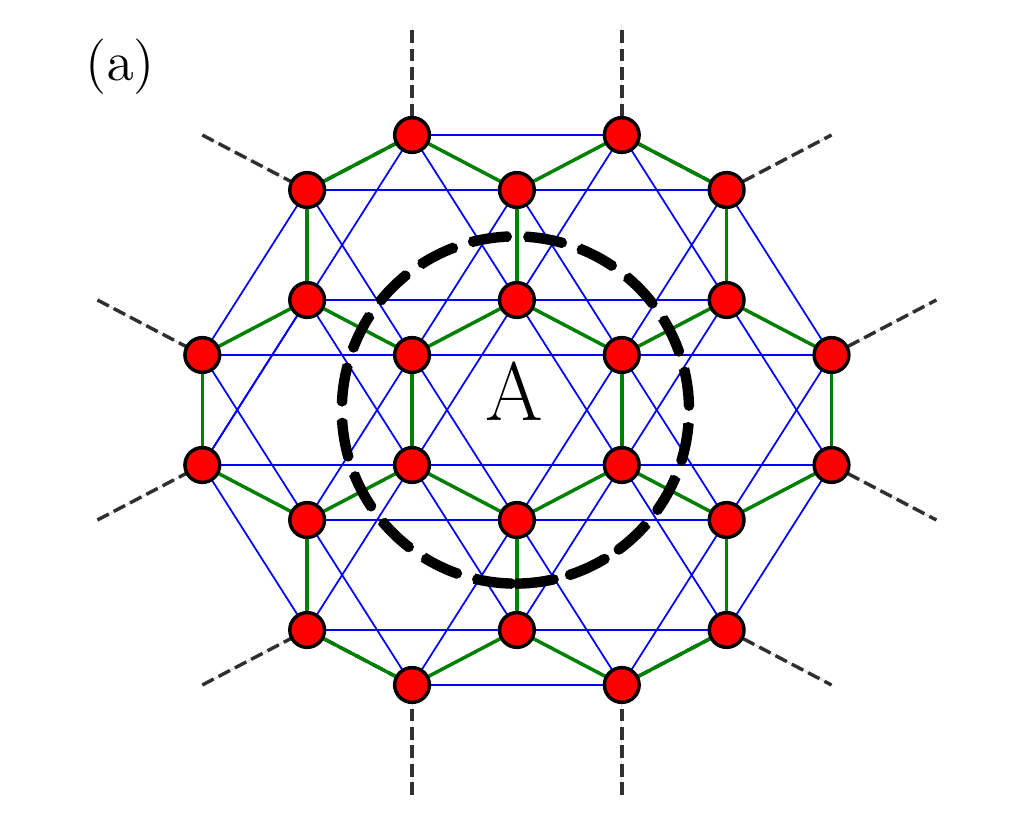,width=1.9in}}
\centerline{\psfig{file=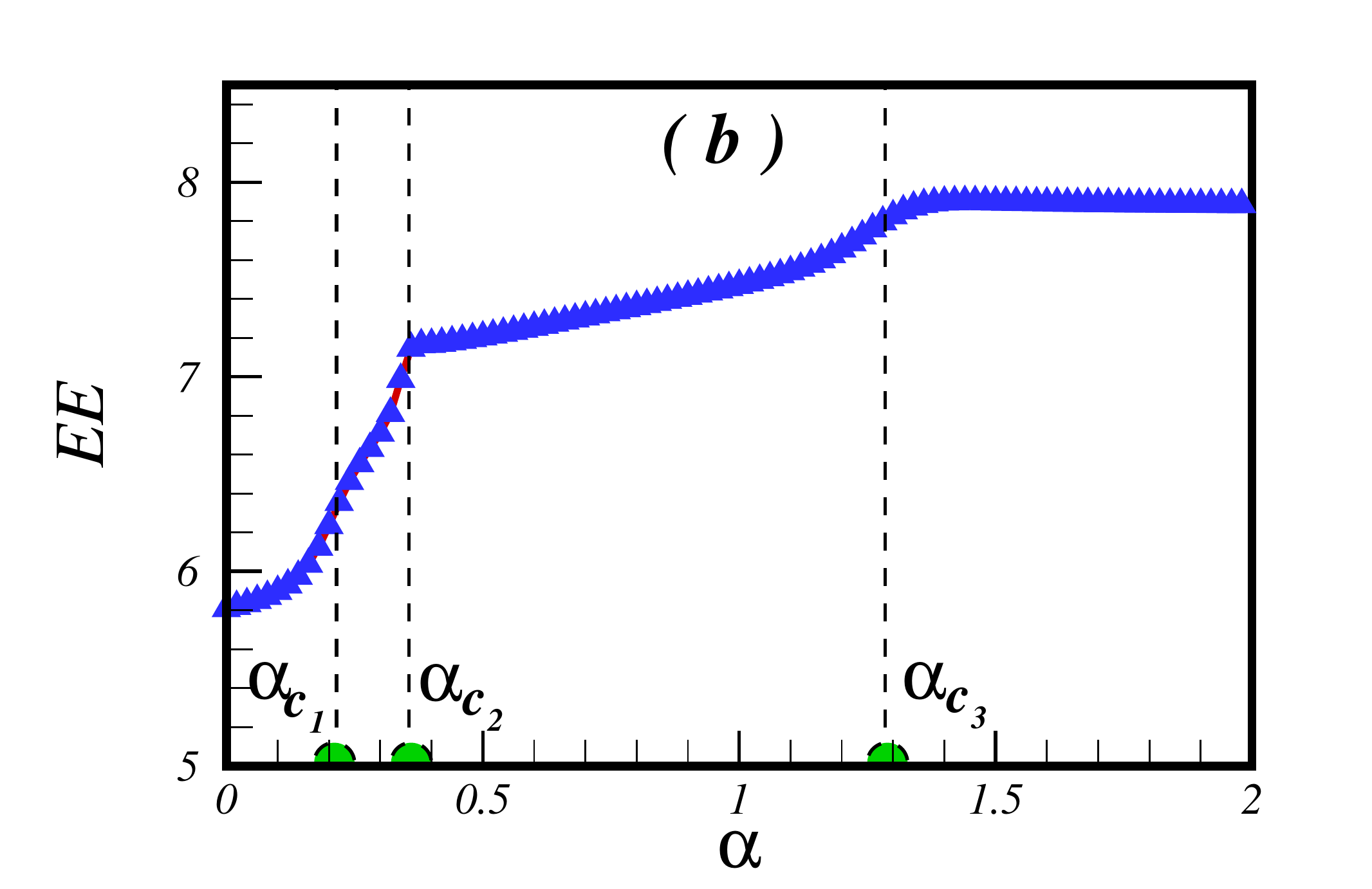,width=1.9in}\psfig{file=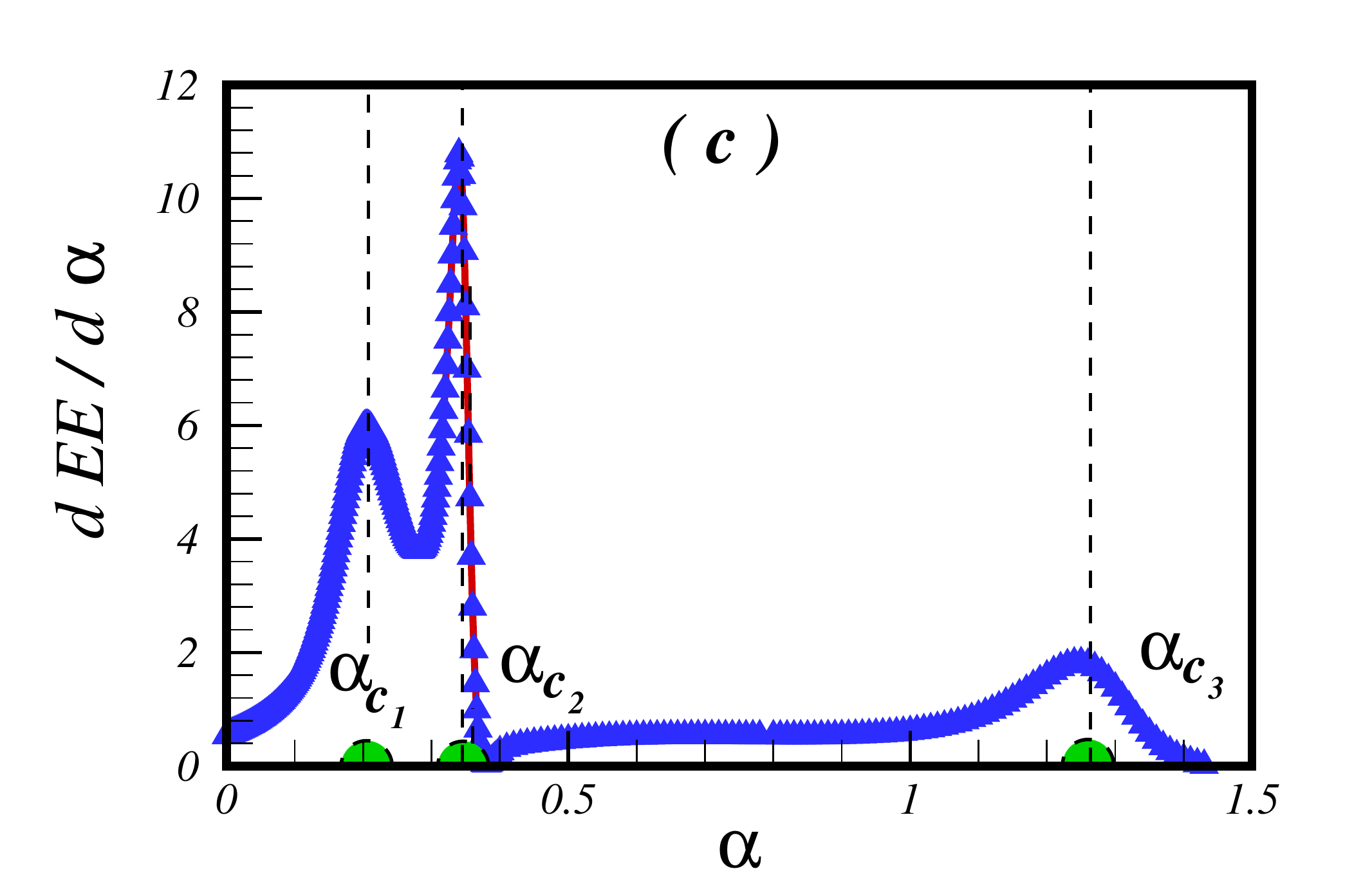,width=1.9in}}
\caption{(a) The subsystem $A$ as a hexagonal cell. (b) The EE versus the frustration parameter $\alpha$ for a cluster of $N=24$ spins. (c) The first derivative of the EE with respect to the frustration. }
\label{fig5}
\end{figure}

Finally, we have calculated the entanglement entropy. To this end, we have split the cluster of $N=24$ spins into two subsystems A and B which subsystem $A$ is selected as an inner hexagonal cell as displayed in Fig.~\ref{fig5}-(a), and the results are presented in Fig.~\ref{fig5}-(b). As is noticed from Fig.~\ref{fig5}-(b), the inner hexagonal cell is entangled with the rest of the lattice in absence of the frustration.  By applying the frustration, the quantum correlation between said hexagonal cell and the rest of the system grows in the N\'eel, the QSL, and the collinear spin-wave phases, although at different rates. Only in the $120^{\circ}$ ordered phase a plateau is observed. Although no direct sign of the first critical point is detected in EE, the location of the second and third critical points are observed. Here, we also calculated the first derivative of EE with respect to the frustration parameter, and the results are plotted in Fig.~\ref{fig5}-(c). Signature of the all critical points is observed in the first derivative of the EE.


\section{CONCLUSION}\label{sec5 }
One interesting challenge in the context of low-dimensional magnets is related to the  spin-$1/2$ anisotropic $XY$ antiferromagnetic Heisenberg honeycomb model. At zero temperature, the model shows the expected long-range N\'eel order.  The big problem is the recognition of induced phases by adding the antiferromagnetic NNN interaction between pairs of spins which is known as frustration.  On the one hand, using the numerical Lanczos method~\cite{FXY1, FXY2}, variational Monte Carlo~\cite{FXY3}, and extended path integral Monte Carlo simulations ~\cite{FXY4}, the QSL phase has been recognized in an intermediate region of frustration. On the other, the existence of the antiferromagnetic Ising phase instead of the quantum spin liquid (QSL) is concluded from the numerical Density matrix renormalization group (DMRG) method~\cite{FXY5, FXY6, FXY11}, and the series expansion methods~\cite{FXY7}.  Since the nature of the system in the intermediate region is coming under doubt, we are motivated with this paper to investigate the ordering of the ground state in the intermediate region indirectly. 
\par
Although in the QSL, phase quantum fluctuations are so dominant as to suppress magnetic ordering at zero temperature, the spin-$1/2$ particles could be entangled~\cite{QSL1, QSL2, QSL3}. On another side, in comparison with the QSL, entanglement between spins in the antiferromagnetic Ising phase should be negligible. For this reason, we focused on the quantum correlations as the concurrence, the quantum discord (QD), and the entanglement of entropy (EE). Using two complementary numerical exact techniques, Lanczos exact diagonalization, and DMRG methods, we calculated numerically the mentioned quantum correlations versus frustration parameter in different clusters.  The nearest neighbor (NN) pair of spins are found entangled in the intermediate region. On the other hand, the next-nearest neighbor (NNN) pairs are not entangled, and frustration can not create entanglement between them. All critical points obtained from the first derivative of the mentioned quantum correlations concerning the frustration are in good agreement with previous results in litterateurs. Moreover, we explicitly establish that the intermediate region is entangled, which support indirectly the existence of  QSL than the Ising phase.

\appendix
\section{Quantum discord}\label{apxa} 
 The QD between a pair of spins  at sites $i$ and $j$ is defined as
\begin{equation}
QD_{i,j} = {\cal I}({\rho _{i,j}}) - {\cal C}({\rho _{i,j}}),
\end{equation}
where the mutual information is given by
\begin{equation}
{\cal I}({\rho _{i,j}}) = S({\rho _{i}}) + S({\rho _{j}}) + \sum\limits_{\alpha  = 0}^3 {{\lambda _\alpha }} \log ({\lambda _\alpha }).
\end{equation}
$\lambda _{\alpha} $ are eigenvalues of $\rho _{i,j}$ and can be read as
\begin{eqnarray} 
\lambda_1 &=& \frac{1}{4}(1+c_3+\sqrt{4c_4^2+(c_1-c_2)^2}),\nonumber\\
\lambda_2 &=& \frac{1}{4}(1+c_3-(\sqrt{4c_4^2+(c_1-c_2)^2}),\nonumber\\
\lambda_3&=&\frac{1}{4}(1-c_3+\vert{c_1+c_2}\vert),\nonumber\\
\lambda_4&=&\frac{1}{4}(1-c_3-\vert{c_1+c_2}\vert),
\label{eigenvalueofrho}
\end{eqnarray}

and 

\begin{equation}
\begin{array}{l}
S({\rho _i}) = S({\rho _{j}}) = \\
\qquad\quad - \left[ {(\frac{{1 + {c_4}}}{2})\log (\frac{{1 + {c_4}}}{2}) + (\frac{{1 - {c_4}}}{2})\log (\frac{{1 - {c_4}}}{2})} \right],
\end{array}
\end{equation}
where we  define new variables as
\begin{eqnarray} 
c_1&=&2 Z_{i,j},\nonumber\\
c_2&=&2 Z_{i,j},\nonumber\\
c_3&=&X_{i,j}^{+}+X_{i,j}^{-}-Y_{i,j}^{+}-Y_{i,j}^{-},\nonumber\\
c_4&=&X_{i,j}^{+}-X_{i,j}^{-}.\nonumber\\
\label{c1c2} 
\end{eqnarray}

The classical correlations, ${\cal C}({\rho _{i,j}})$, is given by

\begin{equation}
\begin{array}{l}
{\cal C}({\rho _{i,j}}) = \\
\mathop {\max }\limits_{\left\{ {\prod\nolimits_i B } \right\}} \left( {S({\rho _i}) - \frac{{S({\rho _0}) + S({\rho _1})}}{2} - {c_4}\cos (\theta )\frac{{S({\rho _0}) - S({\rho _1})}}{2}} \right), 
\end{array}
\end{equation}
where 
\begin{equation}
\begin{array}{l}
S({\rho _k}) =  - \left( {\frac{{1 + {\theta _k}}}{2}} \right)\log \left( {\frac{{1 + {\theta _k}}}{2}} \right) \\
\qquad\qquad - \left( {\frac{{1 - {\theta _k}}}{2}} \right)\log \left( {\frac{{1 - {\theta _k}}}{2}} \right),
\end{array}
\end{equation}
and ${\theta _k} = \sqrt {\sum\limits_{d = 1}^3 {q_{kd}^2} }$. In this equation $q_k$ is defined as
\begin{eqnarray} 
q_{k1}&=&(-1)^{k}
c_1\left[\frac{\sin\theta\cos\phi}{1+(-1)^{k}c_4\cos\theta}\right],\nonumber\\
q_{k2}&=&(-1)^{k}c_2\left[\frac{\sin\theta\sin\phi}{1+(-1)^{k}c_4\cos\theta}\right],\nonumber\\
q_{k3}&=&(-1)^{k}\left[\frac{c_3\cos\theta+(-1)^{k}c_4}{1+(-1)^{k}c_4\cos\theta}\right].
\label{eq27}
\end{eqnarray} 

where $0 \le \theta  \le \pi $ and $0 \le \phi  \le 2\pi $ and $(j)=B$ is a set of projectors for a local measurement on part $j$.

\section{Finite size effect}\label{apxb}

\begin{figure}[t]
\centerline{\psfig{file=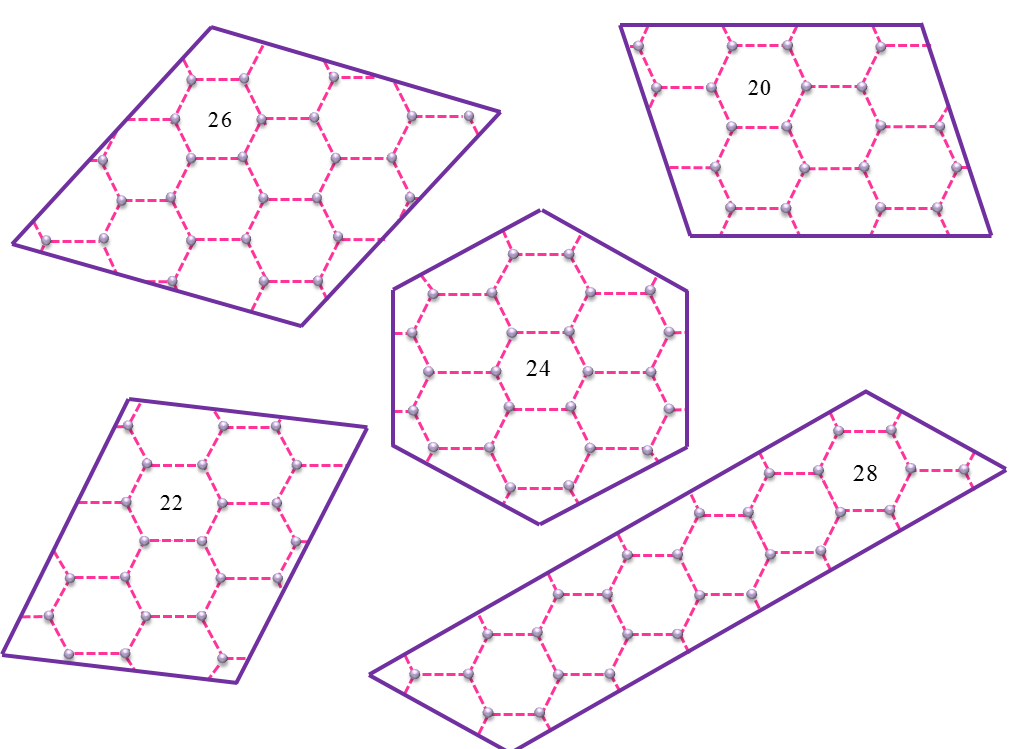,width=3.6in}}
\caption{Clusters used for finite size effect study in Fig.\ref{App2}.}
\label{App1}
\end{figure}

\begin{figure}[t]
\centerline{\psfig{file=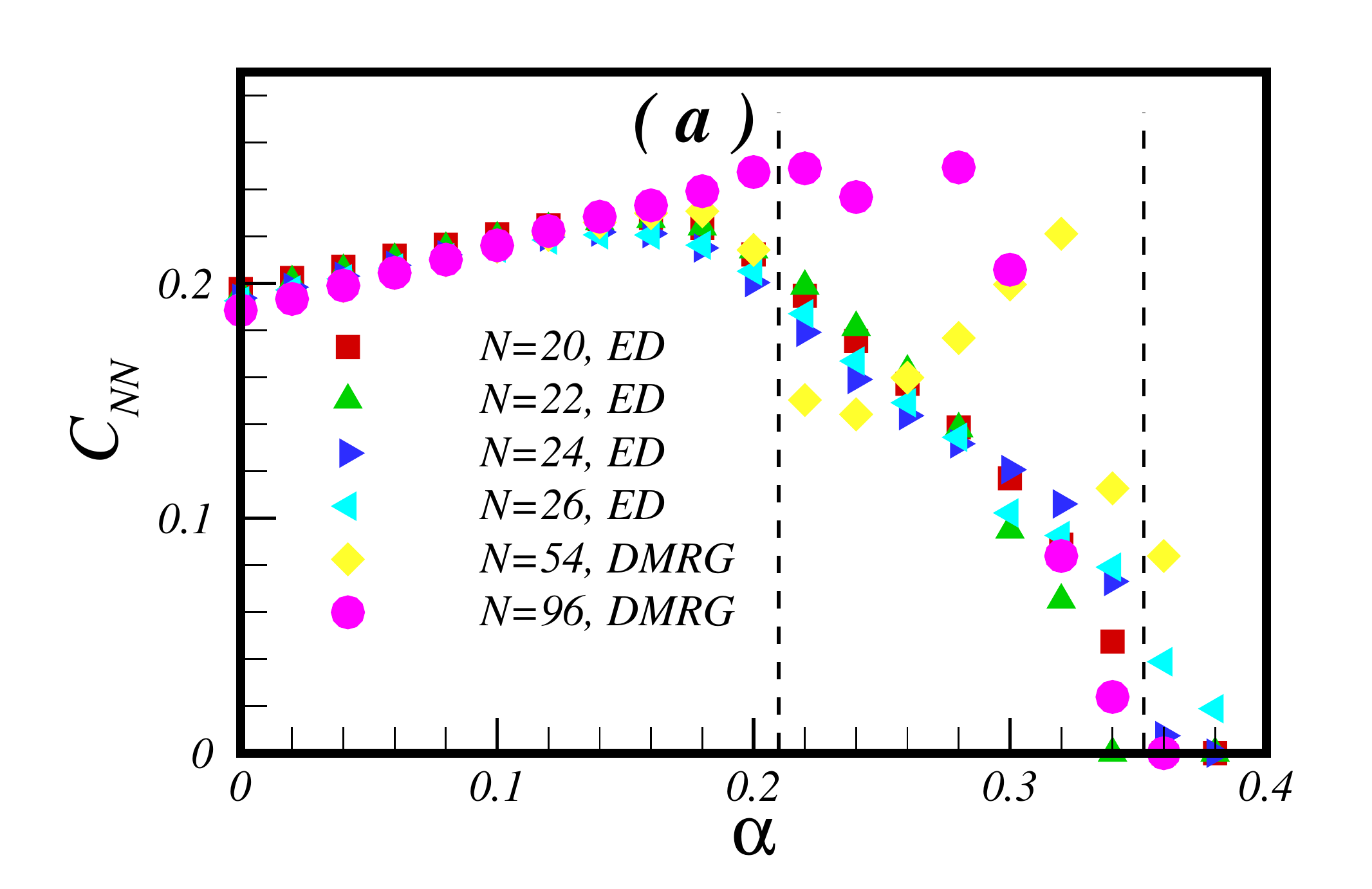,width=1.9in}\psfig{file=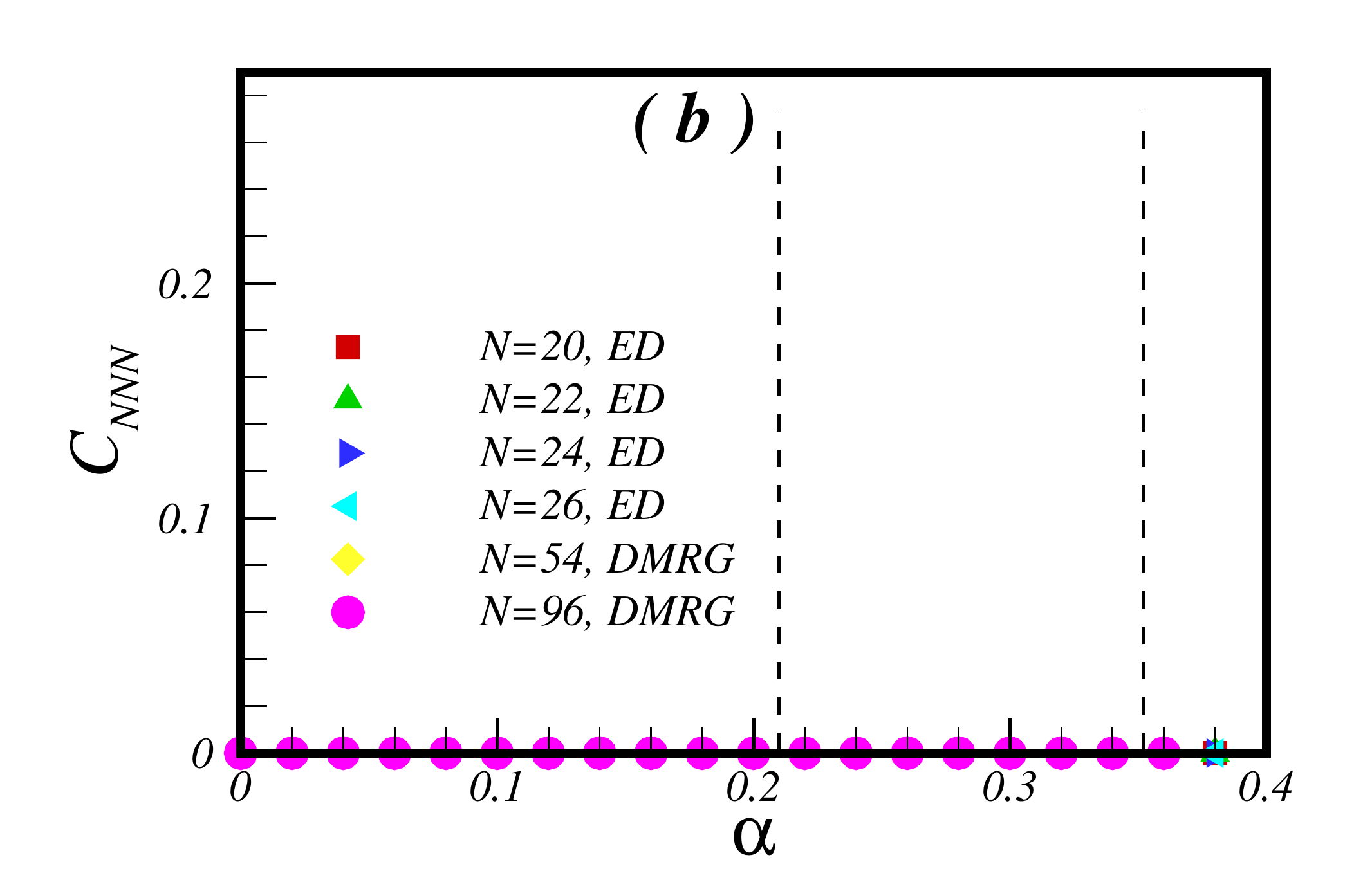,width=1.9in}}
\centerline{\psfig{file=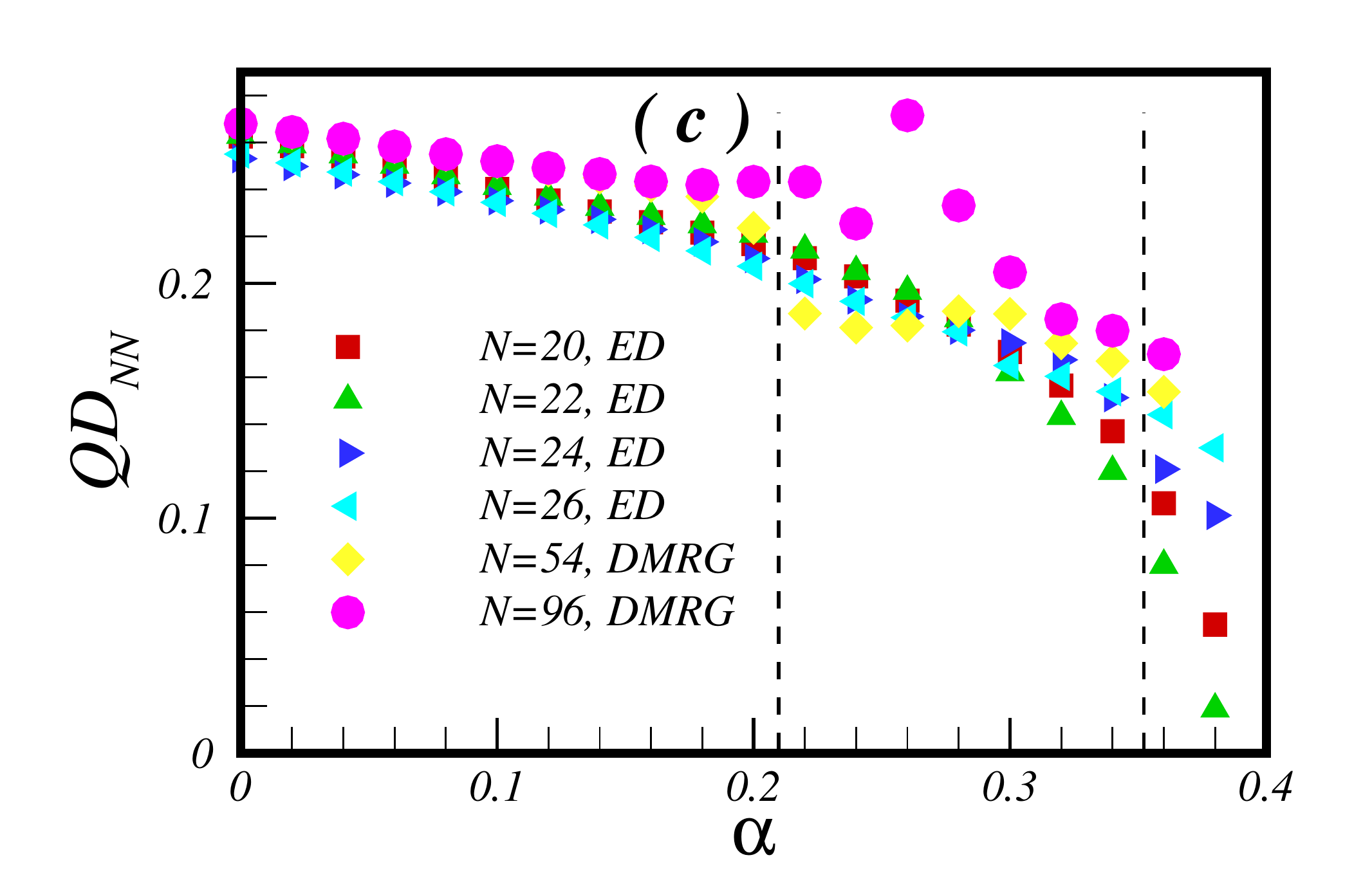,width=1.9in}\psfig{file=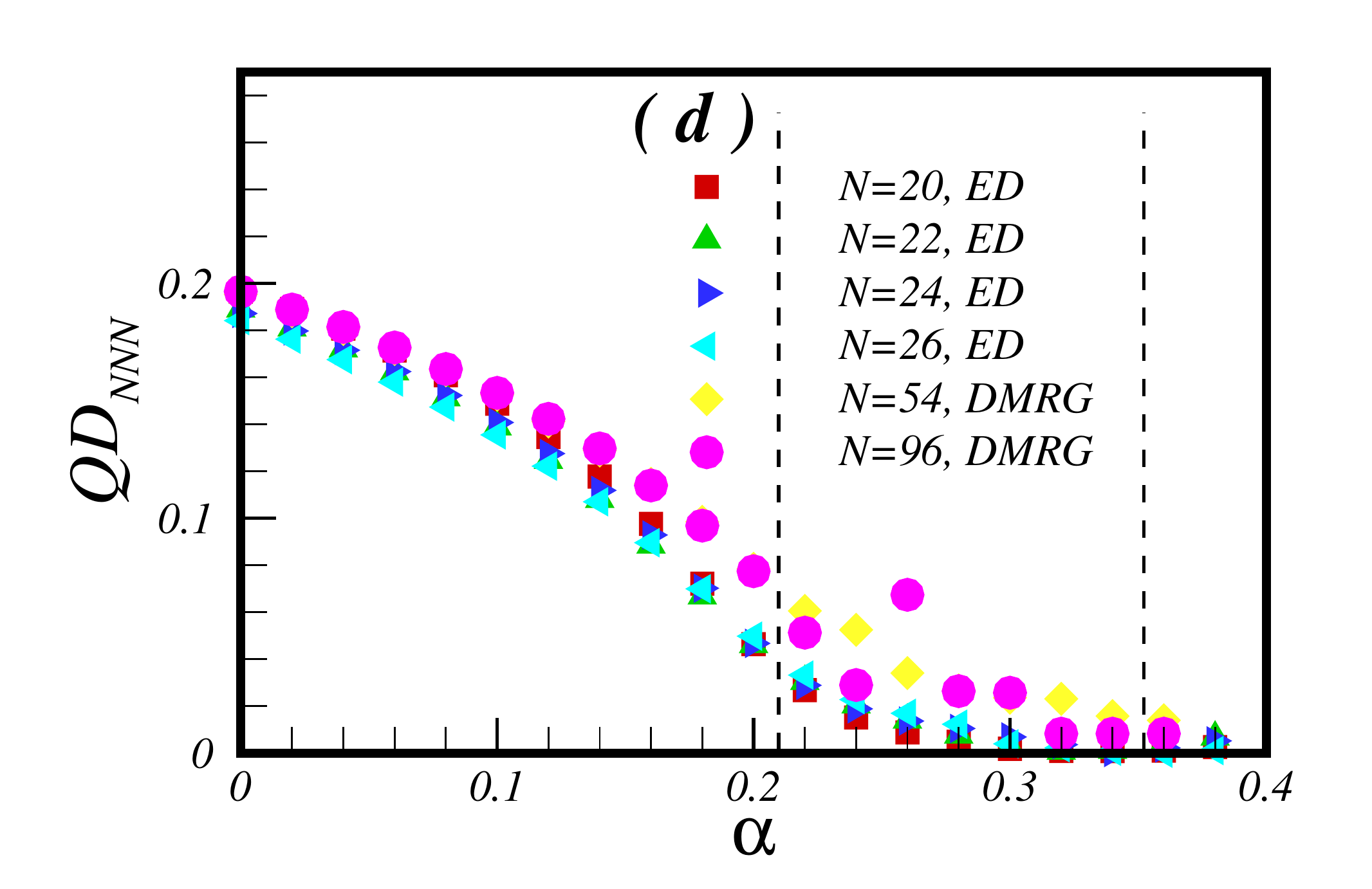,width=1.9in}}
\caption{Finite size study of the concurrence and the QD regrading clusters shown in Fig.\ref{App1}  between the NN ((a) and (c)) and the NNN  ((b) and (d)) pair of spins for  using ED and DMRG methods. Note that for the DMRG we consider the extension of hexagon-shaped cluster shown with $N=24$ in Fig.\ref{App1}.}
\label{App2}
\end{figure}

Here, we present the size effect on the numerical results. Different clusters with $N=20, 22, 24, 26$ are shown in top panel of Fig.~\ref{App1}.  Numerical ED results of the concurrence (Fig.~\ref{App2} (a) and (b)) and the QD (Fig.~\ref{App2} (c) and (d)) are presented. As is seen, extracted  numerical on  all sizes are in good agreement together and show the same behavior with respect to the frustration. We also present  DMRG results for two clusters with size $N=54, 96$, where we consider only the hexagonal geometry. As it can be seen, although DMRG results show some fluctuations in the QSL phase, but show a general trend compare with the ED results.   

%

\section{Entanglement in $120^{\circ}$ ordered phase}\label{apxc}

In the limit of $\alpha \longrightarrow \infty$ where the system is decoupled to $\frac{N}{3}$ independent trimmers constructed of the NNN spins in the honeycomb lattice. The ground state of a trimmer is obtained as   
\begin{eqnarray}
 \left| Gs \right\rangle_{tr} = \frac{1}{\sqrt{2}} [ \left| \uparrow \uparrow \downarrow \right\rangle   -   \left| \uparrow \downarrow \uparrow  \right\rangle    ].
\end{eqnarray}
The reduced density matrix of a pair of spins in the pure $120^{\circ}$ ordered phase is simply find as  
\begin{eqnarray}
\rho_{ij}=\left(
\begin{array}{cccc}
\frac{1}{2} & 0 & 0 & 0 \\
0 & \frac{1}{2} & 0 & 0 \\
0 & 0 & 0 & 0 \\
0 & 0 & 0 & 0 \\
\end{array}
\right),
\end{eqnarray}
which gives zero entanglement between NNN pair of spins. On the other hand, since NN pair of spins are on independent trimmers, the reduced density matrix becomes   

\begin{eqnarray}
\rho_{ij}=\left(
\begin{array}{cccc}
1 & 0 & 0 & 0 \\
0 & 0 & 0 & 0 \\
0 & 0 & 0 & 0 \\
0 & 0 & 0 & 0 \\
\end{array}
\right),
\end{eqnarray}
which shows separability with no entanglement.

\bibliography{Ref/references}

\begin{thebibliography}{70}%
\makeatletter
\providecommand \@ifxundefined [1]{%
 \@ifx{#1\undefined}
}%
\providecommand \@ifnum [1]{%
 \ifnum #1\expandafter \@firstoftwo
 \else \expandafter \@secondoftwo
 \fi
}%
\providecommand \@ifx [1]{%
 \ifx #1\expandafter \@firstoftwo
 \else \expandafter \@secondoftwo
 \fi
}%
\providecommand \natexlab [1]{#1}%
\providecommand \enquote  [1]{``#1''}%
\providecommand \bibnamefont  [1]{#1}%
\providecommand \bibfnamefont [1]{#1}%
\providecommand \citenamefont [1]{#1}%
\providecommand \href@noop [0]{\@secondoftwo}%
\providecommand \href [0]{\begingroup \@sanitize@url \@href}%
\providecommand \@href[1]{\@@startlink{#1}\@@href}%
\providecommand \@@href[1]{\endgroup#1\@@endlink}%
\providecommand \@sanitize@url [0]{\catcode `\\12\catcode `\$12\catcode
  `\&12\catcode `\#12\catcode `\^12\catcode `\_12\catcode `\%12\relax}%
\providecommand \@@startlink[1]{}%
\providecommand \@@endlink[0]{}%
\providecommand \url  [0]{\begingroup\@sanitize@url \@url }%
\providecommand \@url [1]{\endgroup\@href {#1}{\urlprefix }}%
\providecommand \urlprefix  [0]{URL }%
\providecommand \Eprint [0]{\href }%
\providecommand \doibase [0]{https://doi.org/}%
\providecommand \selectlanguage [0]{\@gobble}%
\providecommand \bibinfo  [0]{\@secondoftwo}%
\providecommand \bibfield  [0]{\@secondoftwo}%
\providecommand \translation [1]{[#1]}%
\providecommand \BibitemOpen [0]{}%
\providecommand \bibitemStop [0]{}%
\providecommand \bibitemNoStop [0]{.\EOS\space}%
\providecommand \EOS [0]{\spacefactor3000\relax}%
\providecommand \BibitemShut  [1]{\csname bibitem#1\endcsname}%
\let\auto@bib@innerbib\@empty
\bibitem [{\citenamefont {Sondhi}\ \emph {et~al.}(1997)\citenamefont {Sondhi},
  \citenamefont {Girvin}, \citenamefont {Carini},\ and\ \citenamefont
  {Shahar}}]{a1}%
  \BibitemOpen
  \bibfield  {author} {\bibinfo {author} {\bibfnamefont {S.~L.}\ \bibnamefont
  {Sondhi}}, \bibinfo {author} {\bibfnamefont {S.~M.}\ \bibnamefont {Girvin}},
  \bibinfo {author} {\bibfnamefont {J.~P.}\ \bibnamefont {Carini}},\ and\
  \bibinfo {author} {\bibfnamefont {D.}~\bibnamefont {Shahar}},\ }\bibfield
  {title} {\bibinfo {title} {Continuous quantum phase transitions},\ }\href
  {https://doi.org/10.1103/RevModPhys.69.315} {\bibfield  {journal} {\bibinfo
  {journal} {Rev. Mod. Phys.}\ }\textbf {\bibinfo {volume} {69}},\ \bibinfo
  {pages} {315} (\bibinfo {year} {1997})}\BibitemShut {NoStop}%
\bibitem [{\citenamefont {Sachdev}(1999)}]{a6}%
  \BibitemOpen
  \bibfield  {author} {\bibinfo {author} {\bibfnamefont {S.}~\bibnamefont
  {Sachdev}},\ }\href@noop {} {\emph {\bibinfo {title} {Quantum phase
  transitions}}}\ (\bibinfo  {publisher} {Cambridge University Press},\
  \bibinfo {address} {Cambridge},\ \bibinfo {year} {1999})\BibitemShut
  {NoStop}%
\bibitem [{\citenamefont {C.~Lacroix}(2011)}]{sa1}%
  \BibitemOpen
  \bibfield  {author} {\bibinfo {author} {\bibfnamefont {F.~M.}\ \bibnamefont
  {C.~Lacroix}, \bibfnamefont {P.~Mendels}},\ }\href
  {https://doi.org/https://doi.org/10.1007/978-3-642-10589-0} {\emph {\bibinfo
  {title} {Introduction to Frustrated Magnetism}}}\ (\bibinfo  {publisher}
  {Springer},\ \bibinfo {address} {Berlin, Heidelberg},\ \bibinfo {year}
  {2011})\BibitemShut {NoStop}%
\bibitem [{\citenamefont {Haldane}(1982)}]{sa2}%
  \BibitemOpen
  \bibfield  {author} {\bibinfo {author} {\bibfnamefont {F.~D.~M.}\
  \bibnamefont {Haldane}},\ }\bibfield  {title} {\bibinfo {title} {Spontaneous
  dimerization in the $s=\frac{1}{2}$ heisenberg antiferromagnetic chain with
  competing interactions},\ }\href {https://doi.org/10.1103/PhysRevB.25.4925}
  {\bibfield  {journal} {\bibinfo  {journal} {Phys. Rev. B}\ }\textbf {\bibinfo
  {volume} {25}},\ \bibinfo {pages} {4925} (\bibinfo {year}
  {1982})}\BibitemShut {NoStop}%
\bibitem [{\citenamefont {Okamoto}\ and\ \citenamefont {Nomura}(1992)}]{sa3}%
  \BibitemOpen
  \bibfield  {author} {\bibinfo {author} {\bibfnamefont {K.}~\bibnamefont
  {Okamoto}}\ and\ \bibinfo {author} {\bibfnamefont {K.}~\bibnamefont
  {Nomura}},\ }\bibfield  {title} {\bibinfo {title} {Fluid-dimer critical point
  in s = 12 antiferromagnetic heisenberg chain with next nearest neighbor
  interactions},\ }\href
  {https://doi.org/https://doi.org/10.1016/0375-9601(92)90823-5} {\bibfield
  {journal} {\bibinfo  {journal} {Physics Letters A}\ }\textbf {\bibinfo
  {volume} {169}},\ \bibinfo {pages} {433} (\bibinfo {year}
  {1992})}\BibitemShut {NoStop}%
\bibitem [{\citenamefont {White}\ and\ \citenamefont {Affleck}(1996)}]{sa4}%
  \BibitemOpen
  \bibfield  {author} {\bibinfo {author} {\bibfnamefont {S.~R.}\ \bibnamefont
  {White}}\ and\ \bibinfo {author} {\bibfnamefont {I.}~\bibnamefont
  {Affleck}},\ }\bibfield  {title} {\bibinfo {title} {Dimerization and
  incommensurate spiral spin correlations in the zigzag spin chain: Analogies
  to the kondo lattice},\ }\href {https://doi.org/10.1103/PhysRevB.54.9862}
  {\bibfield  {journal} {\bibinfo  {journal} {Phys. Rev. B}\ }\textbf {\bibinfo
  {volume} {54}},\ \bibinfo {pages} {9862} (\bibinfo {year}
  {1996})}\BibitemShut {NoStop}%
\bibitem [{\citenamefont {G.~Misguich}(2002)}]{sa5}%
  \BibitemOpen
  \bibfield  {author} {\bibinfo {author} {\bibfnamefont {C.~L.}\ \bibnamefont
  {G.~Misguich}},\ }\href
  {https://doi.org/https://doi.org/10.1007/3-540-45649-X_6} {\emph {\bibinfo
  {title} {Frustrated Quantum Magnets}}}\ (\bibinfo  {publisher} {Springer},\
  \bibinfo {address} {Berlin, Heidelberg},\ \bibinfo {year} {2002})\BibitemShut
  {NoStop}%
\bibitem [{\citenamefont {Savary}\ and\ \citenamefont
  {Balents}(2016{\natexlab{a}})}]{QSL1}%
  \BibitemOpen
  \bibfield  {author} {\bibinfo {author} {\bibfnamefont {L.}~\bibnamefont
  {Savary}}\ and\ \bibinfo {author} {\bibfnamefont {L.}~\bibnamefont
  {Balents}},\ }\bibfield  {title} {\bibinfo {title} {Quantum spin liquids: a
  review},\ }\href {https://doi.org/10.1088/0034-4885/80/1/016502} {\bibfield
  {journal} {\bibinfo  {journal} {Reports on Progress in Physics}\ }\textbf
  {\bibinfo {volume} {80}},\ \bibinfo {pages} {016502} (\bibinfo {year}
  {2016}{\natexlab{a}})}\BibitemShut {NoStop}%
\bibitem [{\citenamefont {Zhou}\ \emph {et~al.}(2017)\citenamefont {Zhou},
  \citenamefont {Kanoda},\ and\ \citenamefont {Ng}}]{QSL2}%
  \BibitemOpen
  \bibfield  {author} {\bibinfo {author} {\bibfnamefont {Y.}~\bibnamefont
  {Zhou}}, \bibinfo {author} {\bibfnamefont {K.}~\bibnamefont {Kanoda}},\ and\
  \bibinfo {author} {\bibfnamefont {T.-K.}\ \bibnamefont {Ng}},\ }\bibfield
  {title} {\bibinfo {title} {Quantum spin liquid states},\ }\href
  {https://doi.org/10.1103/RevModPhys.89.025003} {\bibfield  {journal}
  {\bibinfo  {journal} {Rev. Mod. Phys.}\ }\textbf {\bibinfo {volume} {89}},\
  \bibinfo {pages} {025003} (\bibinfo {year} {2017})}\BibitemShut {NoStop}%
\bibitem [{\citenamefont {Knolle}\ and\ \citenamefont {Moessner}(2019)}]{QSL3}%
  \BibitemOpen
  \bibfield  {author} {\bibinfo {author} {\bibfnamefont {J.}~\bibnamefont
  {Knolle}}\ and\ \bibinfo {author} {\bibfnamefont {R.}~\bibnamefont
  {Moessner}},\ }\bibfield  {title} {\bibinfo {title} {A field guide to spin
  liquids},\ }\href {https://doi.org/10.1146/annurev-conmatphys-031218-013401}
  {\bibfield  {journal} {\bibinfo  {journal} {Annual Review of Condensed Matter
  Physics}\ }\textbf {\bibinfo {volume} {10}},\ \bibinfo {pages} {451}
  (\bibinfo {year} {2019})}\BibitemShut {NoStop}%
\bibitem [{\citenamefont {Anderson}(1987)}]{sa7}%
  \BibitemOpen
  \bibfield  {author} {\bibinfo {author} {\bibfnamefont {P.}~\bibnamefont
  {Anderson}},\ }\bibfield  {title} {\bibinfo {title} {The resonating valence
  bond state in la$_2$cuo$_4$ and superconductivity},\ }\href
  {https://doi.org/10.1126/science.235.4793.1196} {\bibfield  {journal}
  {\bibinfo  {journal} {Science}\ }\textbf {\bibinfo {volume} {235}},\ \bibinfo
  {pages} {1196} (\bibinfo {year} {1987})}\BibitemShut {NoStop}%
\bibitem [{\citenamefont {Baskaran}\ \emph {et~al.}(1987)\citenamefont
  {Baskaran}, \citenamefont {Zou},\ and\ \citenamefont {Anderson}}]{sa8}%
  \BibitemOpen
  \bibfield  {author} {\bibinfo {author} {\bibfnamefont {G.}~\bibnamefont
  {Baskaran}}, \bibinfo {author} {\bibfnamefont {Z.}~\bibnamefont {Zou}},\ and\
  \bibinfo {author} {\bibfnamefont {P.}~\bibnamefont {Anderson}},\ }\bibfield
  {title} {\bibinfo {title} {The resonating valence bond state and high-tc
  superconductivity — a mean field theory},\ }\href
  {https://doi.org/https://doi.org/10.1016/0038-1098(87)90642-9} {\bibfield
  {journal} {\bibinfo  {journal} {Solid State Communications}\ }\textbf
  {\bibinfo {volume} {63}},\ \bibinfo {pages} {973} (\bibinfo {year}
  {1987})}\BibitemShut {NoStop}%
\bibitem [{\citenamefont {Anderson}(1973)}]{sa9}%
  \BibitemOpen
  \bibfield  {author} {\bibinfo {author} {\bibfnamefont {P.}~\bibnamefont
  {Anderson}},\ }\bibfield  {title} {\bibinfo {title} {Resonating valence
  bonds: A new kind of insulator?},\ }\href
  {https://doi.org/https://doi.org/10.1016/0025-5408(73)90167-0} {\bibfield
  {journal} {\bibinfo  {journal} {Materials Research Bulletin}\ }\textbf
  {\bibinfo {volume} {8}},\ \bibinfo {pages} {153} (\bibinfo {year}
  {1973})}\BibitemShut {NoStop}%
\bibitem [{\citenamefont {J.~Oitmaa}(1978)}]{H1}%
  \BibitemOpen
  \bibfield  {author} {\bibinfo {author} {\bibfnamefont {D.~D.~B.}\
  \bibnamefont {J.~Oitmaa}},\ }\bibfield  {title} {\bibinfo {title} {The ground
  state of two quantum models of magnetism},\ }\href
  {https://doi.org/10.1139/p78-120} {\bibfield  {journal} {\bibinfo  {journal}
  {Canadian Journal of Physic}\ }\textbf {\bibinfo {volume} {56}} (\bibinfo
  {year} {1978})}\BibitemShut {NoStop}%
\bibitem [{\citenamefont {Reger}\ \emph {et~al.}(1989)\citenamefont {Reger},
  \citenamefont {Riera},\ and\ \citenamefont {Young}}]{H2}%
  \BibitemOpen
  \bibfield  {author} {\bibinfo {author} {\bibfnamefont {J.~D.}\ \bibnamefont
  {Reger}}, \bibinfo {author} {\bibfnamefont {J.~A.}\ \bibnamefont {Riera}},\
  and\ \bibinfo {author} {\bibfnamefont {A.~P.}\ \bibnamefont {Young}},\
  }\bibfield  {title} {\bibinfo {title} {Monte carlo simulations of the
  spin-1/2heisenberg antiferromagnet in two dimensions},\ }\href
  {https://doi.org/10.1088/0953-8984/1/10/007} {\bibfield  {journal} {\bibinfo
  {journal} {Journal of Physics: Condensed Matter}\ }\textbf {\bibinfo {volume}
  {1}},\ \bibinfo {pages} {1855} (\bibinfo {year} {1989})}\BibitemShut
  {NoStop}%
\bibitem [{\citenamefont {Oitmaa}\ \emph {et~al.}(1992)\citenamefont {Oitmaa},
  \citenamefont {Hamer},\ and\ \citenamefont {Weihong}}]{H4}%
  \BibitemOpen
  \bibfield  {author} {\bibinfo {author} {\bibfnamefont {J.}~\bibnamefont
  {Oitmaa}}, \bibinfo {author} {\bibfnamefont {C.~J.}\ \bibnamefont {Hamer}},\
  and\ \bibinfo {author} {\bibfnamefont {Z.}~\bibnamefont {Weihong}},\
  }\bibfield  {title} {\bibinfo {title} {Quantum magnets on the honeycomb and
  triangular lattices at t=0},\ }\href
  {https://doi.org/10.1103/PhysRevB.45.9834} {\bibfield  {journal} {\bibinfo
  {journal} {Phys. Rev. B}\ }\textbf {\bibinfo {volume} {45}},\ \bibinfo
  {pages} {9834} (\bibinfo {year} {1992})}\BibitemShut {NoStop}%
\bibitem [{\citenamefont {{Fouet, J. B.}}\ \emph {et~al.}(2001)\citenamefont
  {{Fouet, J. B.}}, \citenamefont {{Sindzingre, P.}},\ and\ \citenamefont
  {{Lhuillier, C.}}}]{FH1}%
  \BibitemOpen
  \bibfield  {author} {\bibinfo {author} {\bibnamefont {{Fouet, J. B.}}},
  \bibinfo {author} {\bibnamefont {{Sindzingre, P.}}},\ and\ \bibinfo {author}
  {\bibnamefont {{Lhuillier, C.}}},\ }\bibfield  {title} {\bibinfo {title} {An
  investigation of the quantum j1-j2-j3 model on the honeycomb lattice},\
  }\href {https://doi.org/10.1007/s100510170273} {\bibfield  {journal}
  {\bibinfo  {journal} {Eur. Phys. J. B}\ }\textbf {\bibinfo {volume} {20}},\
  \bibinfo {pages} {241} (\bibinfo {year} {2001})}\BibitemShut {NoStop}%
\bibitem [{\citenamefont {Mulder}\ \emph {et~al.}(2010)\citenamefont {Mulder},
  \citenamefont {Ganesh}, \citenamefont {Capriotti},\ and\ \citenamefont
  {Paramekanti}}]{FH2}%
  \BibitemOpen
  \bibfield  {author} {\bibinfo {author} {\bibfnamefont {A.}~\bibnamefont
  {Mulder}}, \bibinfo {author} {\bibfnamefont {R.}~\bibnamefont {Ganesh}},
  \bibinfo {author} {\bibfnamefont {L.}~\bibnamefont {Capriotti}},\ and\
  \bibinfo {author} {\bibfnamefont {A.}~\bibnamefont {Paramekanti}},\
  }\bibfield  {title} {\bibinfo {title} {Spiral order by disorder and lattice
  nematic order in a frustrated heisenberg antiferromagnet on the honeycomb
  lattice},\ }\href {https://doi.org/10.1103/PhysRevB.81.214419} {\bibfield
  {journal} {\bibinfo  {journal} {Phys. Rev. B}\ }\textbf {\bibinfo {volume}
  {81}},\ \bibinfo {pages} {214419} (\bibinfo {year} {2010})}\BibitemShut
  {NoStop}%
\bibitem [{\citenamefont {Meng}\ \emph {et~al.}(2010)\citenamefont {Meng},
  \citenamefont {Lang}, \citenamefont {Wessel}, \citenamefont {Assaad},\ and\
  \citenamefont {Muramatsu}}]{FH3}%
  \BibitemOpen
  \bibfield  {author} {\bibinfo {author} {\bibfnamefont {Z.~Y.}\ \bibnamefont
  {Meng}}, \bibinfo {author} {\bibfnamefont {T.~C.}\ \bibnamefont {Lang}},
  \bibinfo {author} {\bibfnamefont {S.}~\bibnamefont {Wessel}}, \bibinfo
  {author} {\bibfnamefont {F.~F.}\ \bibnamefont {Assaad}},\ and\ \bibinfo
  {author} {\bibfnamefont {A.}~\bibnamefont {Muramatsu}},\ }\bibfield  {title}
  {\bibinfo {title} {Quantum spin liquid emerging in two-dimensional correlated
  dirac fermions},\ }\href {https://doi.org/10.1038/nature08942} {\bibfield
  {journal} {\bibinfo  {journal} {Nature}\ }\textbf {\bibinfo {volume} {464}},\
  \bibinfo {pages} {847} (\bibinfo {year} {2010})}\BibitemShut {NoStop}%
\bibitem [{\citenamefont {Clark}\ \emph {et~al.}(2011)\citenamefont {Clark},
  \citenamefont {Abanin},\ and\ \citenamefont {Sondhi}}]{FH4}%
  \BibitemOpen
  \bibfield  {author} {\bibinfo {author} {\bibfnamefont {B.~K.}\ \bibnamefont
  {Clark}}, \bibinfo {author} {\bibfnamefont {D.~A.}\ \bibnamefont {Abanin}},\
  and\ \bibinfo {author} {\bibfnamefont {S.~L.}\ \bibnamefont {Sondhi}},\
  }\bibfield  {title} {\bibinfo {title} {Nature of the spin liquid state of the
  hubbard model on a honeycomb lattice},\ }\href
  {https://doi.org/10.1103/PhysRevLett.107.087204} {\bibfield  {journal}
  {\bibinfo  {journal} {Phys. Rev. Lett.}\ }\textbf {\bibinfo {volume} {107}},\
  \bibinfo {pages} {087204} (\bibinfo {year} {2011})}\BibitemShut {NoStop}%
\bibitem [{\citenamefont {Albuquerque}\ \emph {et~al.}(2011)\citenamefont
  {Albuquerque}, \citenamefont {Schwandt}, \citenamefont {Het\'enyi},
  \citenamefont {Capponi}, \citenamefont {Mambrini},\ and\ \citenamefont
  {L\"auchli}}]{FH4-1}%
  \BibitemOpen
  \bibfield  {author} {\bibinfo {author} {\bibfnamefont {A.~F.}\ \bibnamefont
  {Albuquerque}}, \bibinfo {author} {\bibfnamefont {D.}~\bibnamefont
  {Schwandt}}, \bibinfo {author} {\bibfnamefont {B.}~\bibnamefont {Het\'enyi}},
  \bibinfo {author} {\bibfnamefont {S.}~\bibnamefont {Capponi}}, \bibinfo
  {author} {\bibfnamefont {M.}~\bibnamefont {Mambrini}},\ and\ \bibinfo
  {author} {\bibfnamefont {A.~M.}\ \bibnamefont {L\"auchli}},\ }\bibfield
  {title} {\bibinfo {title} {Phase diagram of a frustrated quantum
  antiferromagnet on the honeycomb lattice: Magnetic order versus valence-bond
  crystal formation},\ }\href {https://doi.org/10.1103/PhysRevB.84.024406}
  {\bibfield  {journal} {\bibinfo  {journal} {Phys. Rev. B}\ }\textbf {\bibinfo
  {volume} {84}},\ \bibinfo {pages} {024406} (\bibinfo {year}
  {2011})}\BibitemShut {NoStop}%
\bibitem [{\citenamefont {Mosadeq}\ \emph {et~al.}(2011)\citenamefont
  {Mosadeq}, \citenamefont {Shahbazi},\ and\ \citenamefont {Jafari}}]{FH4-2}%
  \BibitemOpen
  \bibfield  {author} {\bibinfo {author} {\bibfnamefont {H.}~\bibnamefont
  {Mosadeq}}, \bibinfo {author} {\bibfnamefont {F.}~\bibnamefont {Shahbazi}},\
  and\ \bibinfo {author} {\bibfnamefont {S.~A.}\ \bibnamefont {Jafari}},\
  }\bibfield  {title} {\bibinfo {title} {Plaquette valence bond ordering in
  {aJ}1{\textendash}j2heisenberg antiferromagnet on a honeycomb lattice},\
  }\href {https://doi.org/10.1088/0953-8984/23/22/226006} {\bibfield  {journal}
  {\bibinfo  {journal} {Journal of Physics: Condensed Matter}\ }\textbf
  {\bibinfo {volume} {23}},\ \bibinfo {pages} {226006} (\bibinfo {year}
  {2011})}\BibitemShut {NoStop}%
\bibitem [{\citenamefont {Oitmaa}\ and\ \citenamefont {Singh}(2011)}]{FH4-3}%
  \BibitemOpen
  \bibfield  {author} {\bibinfo {author} {\bibfnamefont {J.}~\bibnamefont
  {Oitmaa}}\ and\ \bibinfo {author} {\bibfnamefont {R.~R.~P.}\ \bibnamefont
  {Singh}},\ }\bibfield  {title} {\bibinfo {title} {Phase diagram of the
  ${J}_{1}\ensuremath{-}{J}_{2}\ensuremath{-}{J}_{3}$ heisenberg model on the
  honeycomb lattice: A series expansion study},\ }\href
  {https://doi.org/10.1103/PhysRevB.84.094424} {\bibfield  {journal} {\bibinfo
  {journal} {Phys. Rev. B}\ }\textbf {\bibinfo {volume} {84}},\ \bibinfo
  {pages} {094424} (\bibinfo {year} {2011})}\BibitemShut {NoStop}%
\bibitem [{\citenamefont {Mezzacapo}\ and\ \citenamefont
  {Boninsegni}(2012)}]{FH4-4}%
  \BibitemOpen
  \bibfield  {author} {\bibinfo {author} {\bibfnamefont {F.}~\bibnamefont
  {Mezzacapo}}\ and\ \bibinfo {author} {\bibfnamefont {M.}~\bibnamefont
  {Boninsegni}},\ }\bibfield  {title} {\bibinfo {title} {Ground-state phase
  diagram of the quantum ${J}_{1}\ensuremath{-}{J}_{2}$ model on the honeycomb
  lattice},\ }\href {https://doi.org/10.1103/PhysRevB.85.060402} {\bibfield
  {journal} {\bibinfo  {journal} {Phys. Rev. B}\ }\textbf {\bibinfo {volume}
  {85}},\ \bibinfo {pages} {060402} (\bibinfo {year} {2012})}\BibitemShut
  {NoStop}%
\bibitem [{\citenamefont {Bishop}\ \emph {et~al.}(2012)\citenamefont {Bishop},
  \citenamefont {Li}, \citenamefont {Farnell},\ and\ \citenamefont
  {Campbell}}]{FH4-5}%
  \BibitemOpen
  \bibfield  {author} {\bibinfo {author} {\bibfnamefont {R.~F.}\ \bibnamefont
  {Bishop}}, \bibinfo {author} {\bibfnamefont {P.~H.~Y.}\ \bibnamefont {Li}},
  \bibinfo {author} {\bibfnamefont {D.~J.~J.}\ \bibnamefont {Farnell}},\ and\
  \bibinfo {author} {\bibfnamefont {C.~E.}\ \bibnamefont {Campbell}},\
  }\bibfield  {title} {\bibinfo {title} {The frustrated heisenberg
  antiferromagnet on the honeycomb lattice:j1{\textendash}j2model},\ }\href
  {https://doi.org/10.1088/0953-8984/24/23/236002} {\bibfield  {journal}
  {\bibinfo  {journal} {Journal of Physics: Condensed Matter}\ }\textbf
  {\bibinfo {volume} {24}},\ \bibinfo {pages} {236002} (\bibinfo {year}
  {2012})}\BibitemShut {NoStop}%
\bibitem [{\citenamefont {Smirnova}\ \emph {et~al.}(2009)\citenamefont
  {Smirnova}, \citenamefont {Azuma}, \citenamefont {Kumada}, \citenamefont
  {Kusano}, \citenamefont {Matsuda}, \citenamefont {Shimakawa}, \citenamefont
  {Takei}, \citenamefont {Yonesaki},\ and\ \citenamefont {Kinomura}}]{FH5}%
  \BibitemOpen
  \bibfield  {author} {\bibinfo {author} {\bibfnamefont {O.}~\bibnamefont
  {Smirnova}}, \bibinfo {author} {\bibfnamefont {M.}~\bibnamefont {Azuma}},
  \bibinfo {author} {\bibfnamefont {N.}~\bibnamefont {Kumada}}, \bibinfo
  {author} {\bibfnamefont {Y.}~\bibnamefont {Kusano}}, \bibinfo {author}
  {\bibfnamefont {M.}~\bibnamefont {Matsuda}}, \bibinfo {author} {\bibfnamefont
  {Y.}~\bibnamefont {Shimakawa}}, \bibinfo {author} {\bibfnamefont
  {T.}~\bibnamefont {Takei}}, \bibinfo {author} {\bibfnamefont
  {Y.}~\bibnamefont {Yonesaki}},\ and\ \bibinfo {author} {\bibfnamefont
  {N.}~\bibnamefont {Kinomura}},\ }\bibfield  {title} {\bibinfo {title}
  {Synthesis, crystal structure, and magnetic properties of bi3mn4o12(no3)
  oxynitrate comprising s = 3/2 honeycomb lattice},\ }\href
  {https://doi.org/10.1021/ja901922p} {\bibfield  {journal} {\bibinfo
  {journal} {Journal of the American Chemical Society}\ }\textbf {\bibinfo
  {volume} {131}},\ \bibinfo {pages} {8313} (\bibinfo {year}
  {2009})}\BibitemShut {NoStop}%
\bibitem [{\citenamefont {Okubo}\ \emph {et~al.}(2010)\citenamefont {Okubo},
  \citenamefont {Elmasry}, \citenamefont {Zhang}, \citenamefont {Fujisawa},
  \citenamefont {Sakurai}, \citenamefont {Ohta}, \citenamefont {Azuma},
  \citenamefont {Sumirnova},\ and\ \citenamefont {Kumada}}]{FH6}%
  \BibitemOpen
  \bibfield  {author} {\bibinfo {author} {\bibfnamefont {.}~\bibnamefont
  {Okubo}}, \bibinfo {author} {\bibfnamefont {F.}~\bibnamefont {Elmasry}},
  \bibinfo {author} {\bibfnamefont {.}~\bibnamefont {Zhang}}, \bibinfo {author}
  {\bibfnamefont {M.}~\bibnamefont {Fujisawa}}, \bibinfo {author}
  {\bibfnamefont {T.}~\bibnamefont {Sakurai}}, \bibinfo {author} {\bibfnamefont
  {H.}~\bibnamefont {Ohta}}, \bibinfo {author} {\bibfnamefont {M.}~\bibnamefont
  {Azuma}}, \bibinfo {author} {\bibfnamefont {O.~A.}\ \bibnamefont
  {Sumirnova}},\ and\ \bibinfo {author} {\bibfnamefont {N.}~\bibnamefont
  {Kumada}},\ }\bibfield  {title} {\bibinfo {title} {High-field {ESR}
  measurements {ofS}=3/2 honeycomb lattice antiferromagnet bi3mn4o12({NO}3)},\
  }\href {https://doi.org/10.1088/1742-6596/200/2/022042} {\bibfield  {journal}
  {\bibinfo  {journal} {Journal of Physics: Conference Series}\ }\textbf
  {\bibinfo {volume} {200}},\ \bibinfo {pages} {022042} (\bibinfo {year}
  {2010})}\BibitemShut {NoStop}%
\bibitem [{\citenamefont {Varney}\ \emph {et~al.}(2011)\citenamefont {Varney},
  \citenamefont {Sun}, \citenamefont {Galitski},\ and\ \citenamefont
  {Rigol}}]{FXY1}%
  \BibitemOpen
  \bibfield  {author} {\bibinfo {author} {\bibfnamefont {C.~N.}\ \bibnamefont
  {Varney}}, \bibinfo {author} {\bibfnamefont {K.}~\bibnamefont {Sun}},
  \bibinfo {author} {\bibfnamefont {V.}~\bibnamefont {Galitski}},\ and\
  \bibinfo {author} {\bibfnamefont {M.}~\bibnamefont {Rigol}},\ }\bibfield
  {title} {\bibinfo {title} {Kaleidoscope of exotic quantum phases in a
  frustrated $xy$ model},\ }\href
  {https://doi.org/10.1103/PhysRevLett.107.077201} {\bibfield  {journal}
  {\bibinfo  {journal} {Phys. Rev. Lett.}\ }\textbf {\bibinfo {volume} {107}},\
  \bibinfo {pages} {077201} (\bibinfo {year} {2011})}\BibitemShut {NoStop}%
\bibitem [{\citenamefont {Varney}\ \emph {et~al.}(2012)\citenamefont {Varney},
  \citenamefont {Sun}, \citenamefont {Galitski},\ and\ \citenamefont
  {Rigol}}]{FXY2}%
  \BibitemOpen
  \bibfield  {author} {\bibinfo {author} {\bibfnamefont {C.~N.}\ \bibnamefont
  {Varney}}, \bibinfo {author} {\bibfnamefont {K.}~\bibnamefont {Sun}},
  \bibinfo {author} {\bibfnamefont {V.}~\bibnamefont {Galitski}},\ and\
  \bibinfo {author} {\bibfnamefont {M.}~\bibnamefont {Rigol}},\ }\bibfield
  {title} {\bibinfo {title} {Quantum phases of hard-core bosons in a frustrated
  honeycomb lattice},\ }\href {https://doi.org/10.1088/1367-2630/14/11/115028}
  {\bibfield  {journal} {\bibinfo  {journal} {New Journal of Physics}\ }\textbf
  {\bibinfo {volume} {14}},\ \bibinfo {pages} {115028} (\bibinfo {year}
  {2012})}\BibitemShut {NoStop}%
\bibitem [{\citenamefont {Carrasquilla}\ \emph {et~al.}(2013)\citenamefont
  {Carrasquilla}, \citenamefont {Ciolo}, \citenamefont {Becca}, \citenamefont
  {Galitski},\ and\ \citenamefont {Rigol}}]{FXY3}%
  \BibitemOpen
  \bibfield  {author} {\bibinfo {author} {\bibfnamefont {J.}~\bibnamefont
  {Carrasquilla}}, \bibinfo {author} {\bibfnamefont {A.~D.}\ \bibnamefont
  {Ciolo}}, \bibinfo {author} {\bibfnamefont {F.}~\bibnamefont {Becca}},
  \bibinfo {author} {\bibfnamefont {V.}~\bibnamefont {Galitski}},\ and\
  \bibinfo {author} {\bibfnamefont {M.}~\bibnamefont {Rigol}},\ }\bibfield
  {title} {\bibinfo {title} {Nature of the phases in the frustrated $xy$ model
  on the honeycomb lattice},\ }\href
  {https://doi.org/10.1103/PhysRevB.88.241109} {\bibfield  {journal} {\bibinfo
  {journal} {Phys. Rev. B}\ }\textbf {\bibinfo {volume} {88}},\ \bibinfo
  {pages} {241109} (\bibinfo {year} {2013})}\BibitemShut {NoStop}%
\bibitem [{\citenamefont {Nakafuji}\ and\ \citenamefont
  {Ichinose}(2017)}]{FXY8}%
  \BibitemOpen
  \bibfield  {author} {\bibinfo {author} {\bibfnamefont {T.}~\bibnamefont
  {Nakafuji}}\ and\ \bibinfo {author} {\bibfnamefont {I.}~\bibnamefont
  {Ichinose}},\ }\bibfield  {title} {\bibinfo {title} {Phase diagrams of
  bose-hubbard model and antiferromagnetic spin-1/2 models on a honeycomb
  lattice},\ }\href {https://doi.org/10.1103/PhysRevA.96.013628} {\bibfield
  {journal} {\bibinfo  {journal} {Phys. Rev. A}\ }\textbf {\bibinfo {volume}
  {96}},\ \bibinfo {pages} {013628} (\bibinfo {year} {2017})}\BibitemShut
  {NoStop}%
\bibitem [{\citenamefont {Di~Ciolo}\ \emph {et~al.}(2014)\citenamefont
  {Di~Ciolo}, \citenamefont {Carrasquilla}, \citenamefont {Becca},
  \citenamefont {Rigol},\ and\ \citenamefont {Galitski}}]{FXY4}%
  \BibitemOpen
  \bibfield  {author} {\bibinfo {author} {\bibfnamefont {A.}~\bibnamefont
  {Di~Ciolo}}, \bibinfo {author} {\bibfnamefont {J.}~\bibnamefont
  {Carrasquilla}}, \bibinfo {author} {\bibfnamefont {F.}~\bibnamefont {Becca}},
  \bibinfo {author} {\bibfnamefont {M.}~\bibnamefont {Rigol}},\ and\ \bibinfo
  {author} {\bibfnamefont {V.}~\bibnamefont {Galitski}},\ }\bibfield  {title}
  {\bibinfo {title} {Spiral antiferromagnets beyond the spin-wave
  approximation: Frustrated $xy$ and heisenberg models on the honeycomb
  lattice},\ }\href {https://doi.org/10.1103/PhysRevB.89.094413} {\bibfield
  {journal} {\bibinfo  {journal} {Phys. Rev. B}\ }\textbf {\bibinfo {volume}
  {89}},\ \bibinfo {pages} {094413} (\bibinfo {year} {2014})}\BibitemShut
  {NoStop}%
\bibitem [{\citenamefont {Zhu}\ \emph {et~al.}(2013)\citenamefont {Zhu},
  \citenamefont {Huse},\ and\ \citenamefont {White}}]{FXY5}%
  \BibitemOpen
  \bibfield  {author} {\bibinfo {author} {\bibfnamefont {Z.}~\bibnamefont
  {Zhu}}, \bibinfo {author} {\bibfnamefont {D.~A.}\ \bibnamefont {Huse}},\ and\
  \bibinfo {author} {\bibfnamefont {S.~R.}\ \bibnamefont {White}},\ }\bibfield
  {title} {\bibinfo {title} {Unexpected $z$-direction ising antiferromagnetic
  order in a frustrated spin-$1/2$ ${J}_{1}\ensuremath{-}{J}_{2}$ $xy$ model on
  the honeycomb lattice},\ }\href
  {https://doi.org/10.1103/PhysRevLett.111.257201} {\bibfield  {journal}
  {\bibinfo  {journal} {Phys. Rev. Lett.}\ }\textbf {\bibinfo {volume} {111}},\
  \bibinfo {pages} {257201} (\bibinfo {year} {2013})}\BibitemShut {NoStop}%
\bibitem [{\citenamefont {Zhu}\ and\ \citenamefont {White}(2014)}]{FXY6}%
  \BibitemOpen
  \bibfield  {author} {\bibinfo {author} {\bibfnamefont {Z.}~\bibnamefont
  {Zhu}}\ and\ \bibinfo {author} {\bibfnamefont {S.~R.}\ \bibnamefont
  {White}},\ }\bibfield  {title} {\bibinfo {title} {Quantum phases of the
  frustrated xy models on the honeycomb lattice},\ }\href
  {https://doi.org/10.1142/S0217984914300166} {\bibfield  {journal} {\bibinfo
  {journal} {Modern Physics Letters B}\ }\textbf {\bibinfo {volume} {28}},\
  \bibinfo {pages} {1430016} (\bibinfo {year} {2014})}\BibitemShut {NoStop}%
\bibitem [{\citenamefont {Oitmaa}\ and\ \citenamefont {Singh}(2014)}]{FXY7}%
  \BibitemOpen
  \bibfield  {author} {\bibinfo {author} {\bibfnamefont {J.}~\bibnamefont
  {Oitmaa}}\ and\ \bibinfo {author} {\bibfnamefont {R.~R.~P.}\ \bibnamefont
  {Singh}},\ }\bibfield  {title} {\bibinfo {title} {Phase diagram of the
  frustrated quantum-$xy$ model on the honeycomb lattice studied by series
  expansions: Evidence for proximity to a bicritical point},\ }\href
  {https://doi.org/10.1103/PhysRevB.89.104423} {\bibfield  {journal} {\bibinfo
  {journal} {Phys. Rev. B}\ }\textbf {\bibinfo {volume} {89}},\ \bibinfo
  {pages} {104423} (\bibinfo {year} {2014})}\BibitemShut {NoStop}%
\bibitem [{\citenamefont {Huang}\ \emph {et~al.}(2021)\citenamefont {Huang},
  \citenamefont {Dong}, \citenamefont {Sheng},\ and\ \citenamefont
  {Ting}}]{FXY11}%
  \BibitemOpen
  \bibfield  {author} {\bibinfo {author} {\bibfnamefont {Y.}~\bibnamefont
  {Huang}}, \bibinfo {author} {\bibfnamefont {X.-Y.}\ \bibnamefont {Dong}},
  \bibinfo {author} {\bibfnamefont {D.~N.}\ \bibnamefont {Sheng}},\ and\
  \bibinfo {author} {\bibfnamefont {C.~S.}\ \bibnamefont {Ting}},\ }\bibfield
  {title} {\bibinfo {title} {Quantum phase diagram and chiral spin liquid in
  the extended spin-$\frac{1}{2}$ honeycomb xy model},\ }\href
  {https://doi.org/10.1103/PhysRevB.103.L041108} {\bibfield  {journal}
  {\bibinfo  {journal} {Phys. Rev. B}\ }\textbf {\bibinfo {volume} {103}},\
  \bibinfo {pages} {L041108} (\bibinfo {year} {2021})}\BibitemShut {NoStop}%
\bibitem [{\citenamefont {Ma}(2018)}]{FXY9}%
  \BibitemOpen
  \bibfield  {author} {\bibinfo {author} {\bibfnamefont {H.}~\bibnamefont
  {Ma}},\ }\bibfield  {title} {\bibinfo {title} {Possible phases of the
  spin-$\frac{1}{2}$ xxz model on a honeycomb lattice by boson-vortex
  duality},\ }\href {https://doi.org/10.1103/PhysRevB.97.045104} {\bibfield
  {journal} {\bibinfo  {journal} {Phys. Rev. B}\ }\textbf {\bibinfo {volume}
  {97}},\ \bibinfo {pages} {045104} (\bibinfo {year} {2018})}\BibitemShut
  {NoStop}%
\bibitem [{\citenamefont {Plekhanov}\ \emph {et~al.}(2018)\citenamefont
  {Plekhanov}, \citenamefont {Vasi\ifmmode~\acute{c}\else \'{c}\fi{}},
  \citenamefont {Petrescu}, \citenamefont {Nirwan}, \citenamefont {Roux},
  \citenamefont {Hofstetter},\ and\ \citenamefont {Le~Hur}}]{FXY10}%
  \BibitemOpen
  \bibfield  {author} {\bibinfo {author} {\bibfnamefont {K.}~\bibnamefont
  {Plekhanov}}, \bibinfo {author} {\bibfnamefont {I.}~\bibnamefont
  {Vasi\ifmmode~\acute{c}\else \'{c}\fi{}}}, \bibinfo {author} {\bibfnamefont
  {A.}~\bibnamefont {Petrescu}}, \bibinfo {author} {\bibfnamefont
  {R.}~\bibnamefont {Nirwan}}, \bibinfo {author} {\bibfnamefont
  {G.}~\bibnamefont {Roux}}, \bibinfo {author} {\bibfnamefont {W.}~\bibnamefont
  {Hofstetter}},\ and\ \bibinfo {author} {\bibfnamefont {K.}~\bibnamefont
  {Le~Hur}},\ }\bibfield  {title} {\bibinfo {title} {Emergent chiral spin state
  in the mott phase of a bosonic kane-mele-hubbard model},\ }\href
  {https://doi.org/10.1103/PhysRevLett.120.157201} {\bibfield  {journal}
  {\bibinfo  {journal} {Phys. Rev. Lett.}\ }\textbf {\bibinfo {volume} {120}},\
  \bibinfo {pages} {157201} (\bibinfo {year} {2018})}\BibitemShut {NoStop}%
\bibitem [{\citenamefont {R.~Wang}(2010)}]{FXY12}%
  \BibitemOpen
  \bibfield  {author} {\bibinfo {author} {\bibfnamefont {T.~A.~S.}\
  \bibnamefont {R.~Wang}, \bibfnamefont {B.~Wang}},\ }\bibfield  {title}
  {\bibinfo {title} {Chern-simons superconductors and their instabilities},\
  }\href {https://arxiv.org/abs/2010.10067} {\bibfield  {journal} {\bibinfo
  {journal} {arXiv:2010.10067}\ } (\bibinfo {year} {2010})}\BibitemShut
  {NoStop}%
\bibitem [{\citenamefont {Satoori}\ \emph {et~al.}(2022)\citenamefont
  {Satoori}, \citenamefont {Mahdavifar},\ and\ \citenamefont
  {Vahedi}}]{Satoori_2022}%
  \BibitemOpen
  \bibfield  {author} {\bibinfo {author} {\bibfnamefont {S.}~\bibnamefont
  {Satoori}}, \bibinfo {author} {\bibfnamefont {S.}~\bibnamefont
  {Mahdavifar}},\ and\ \bibinfo {author} {\bibfnamefont {J.}~\bibnamefont
  {Vahedi}},\ }\href@noop {} {\bibinfo {title} {Entanglement and quantum
  correlations in the xx spin-$1/2$ honeycomb lattice}} (\bibinfo {year}
  {2022}),\ \Eprint {https://arxiv.org/abs/arXiv:2204.07708} {arXiv:2204.07708}
  \BibitemShut {NoStop}%
\bibitem [{\citenamefont {Amico}\ \emph {et~al.}(2008)\citenamefont {Amico},
  \citenamefont {Fazio}, \citenamefont {Osterloh},\ and\ \citenamefont
  {Vedral}}]{E1}%
  \BibitemOpen
  \bibfield  {author} {\bibinfo {author} {\bibfnamefont {L.}~\bibnamefont
  {Amico}}, \bibinfo {author} {\bibfnamefont {R.}~\bibnamefont {Fazio}},
  \bibinfo {author} {\bibfnamefont {A.}~\bibnamefont {Osterloh}},\ and\
  \bibinfo {author} {\bibfnamefont {V.}~\bibnamefont {Vedral}},\ }\bibfield
  {title} {\bibinfo {title} {Entanglement in many-body systems},\ }\href
  {https://doi.org/10.1103/RevModPhys.80.517} {\bibfield  {journal} {\bibinfo
  {journal} {Rev. Mod. Phys.}\ }\textbf {\bibinfo {volume} {80}},\ \bibinfo
  {pages} {517} (\bibinfo {year} {2008})}\BibitemShut {NoStop}%
\bibitem [{\citenamefont {Horodecki}\ \emph {et~al.}(2009)\citenamefont
  {Horodecki}, \citenamefont {Horodecki}, \citenamefont {Horodecki},\ and\
  \citenamefont {Horodecki}}]{E2}%
  \BibitemOpen
  \bibfield  {author} {\bibinfo {author} {\bibfnamefont {R.}~\bibnamefont
  {Horodecki}}, \bibinfo {author} {\bibfnamefont {P.}~\bibnamefont
  {Horodecki}}, \bibinfo {author} {\bibfnamefont {M.}~\bibnamefont
  {Horodecki}},\ and\ \bibinfo {author} {\bibfnamefont {K.}~\bibnamefont
  {Horodecki}},\ }\bibfield  {title} {\bibinfo {title} {Quantum entanglement},\
  }\href {https://doi.org/10.1103/RevModPhys.81.865} {\bibfield  {journal}
  {\bibinfo  {journal} {Rev. Mod. Phys.}\ }\textbf {\bibinfo {volume} {81}},\
  \bibinfo {pages} {865} (\bibinfo {year} {2009})}\BibitemShut {NoStop}%
\bibitem [{\citenamefont {Bera}\ \emph {et~al.}(2017)\citenamefont {Bera},
  \citenamefont {Das}, \citenamefont {Sadhukhan}, \citenamefont {Roy},
  \citenamefont {Sen(De)},\ and\ \citenamefont {Sen}}]{E3}%
  \BibitemOpen
  \bibfield  {author} {\bibinfo {author} {\bibfnamefont {A.}~\bibnamefont
  {Bera}}, \bibinfo {author} {\bibfnamefont {T.}~\bibnamefont {Das}}, \bibinfo
  {author} {\bibfnamefont {D.}~\bibnamefont {Sadhukhan}}, \bibinfo {author}
  {\bibfnamefont {S.~S.}\ \bibnamefont {Roy}}, \bibinfo {author} {\bibfnamefont
  {A.}~\bibnamefont {Sen(De)}},\ and\ \bibinfo {author} {\bibfnamefont
  {U.}~\bibnamefont {Sen}},\ }\bibfield  {title} {\bibinfo {title} {Quantum
  discord and its allies: a review of recent progress},\ }\href
  {https://doi.org/10.1088/1361-6633/aa872f} {\bibfield  {journal} {\bibinfo
  {journal} {Reports on Progress in Physics}\ }\textbf {\bibinfo {volume}
  {81}},\ \bibinfo {pages} {024001} (\bibinfo {year} {2017})}\BibitemShut
  {NoStop}%
\bibitem [{\citenamefont {Braun}\ \emph {et~al.}(2018)\citenamefont {Braun},
  \citenamefont {Adesso}, \citenamefont {Benatti}, \citenamefont {Floreanini},
  \citenamefont {Marzolino}, \citenamefont {Mitchell},\ and\ \citenamefont
  {Pirandola}}]{E4}%
  \BibitemOpen
  \bibfield  {author} {\bibinfo {author} {\bibfnamefont {D.}~\bibnamefont
  {Braun}}, \bibinfo {author} {\bibfnamefont {G.}~\bibnamefont {Adesso}},
  \bibinfo {author} {\bibfnamefont {F.}~\bibnamefont {Benatti}}, \bibinfo
  {author} {\bibfnamefont {R.}~\bibnamefont {Floreanini}}, \bibinfo {author}
  {\bibfnamefont {U.}~\bibnamefont {Marzolino}}, \bibinfo {author}
  {\bibfnamefont {M.~W.}\ \bibnamefont {Mitchell}},\ and\ \bibinfo {author}
  {\bibfnamefont {S.}~\bibnamefont {Pirandola}},\ }\bibfield  {title} {\bibinfo
  {title} {Quantum-enhanced measurements without entanglement},\ }\href
  {https://doi.org/10.1103/RevModPhys.90.035006} {\bibfield  {journal}
  {\bibinfo  {journal} {Rev. Mod. Phys.}\ }\textbf {\bibinfo {volume} {90}},\
  \bibinfo {pages} {035006} (\bibinfo {year} {2018})}\BibitemShut {NoStop}%
\bibitem [{\citenamefont {Balents}(2010)}]{6}%
  \BibitemOpen
  \bibfield  {author} {\bibinfo {author} {\bibfnamefont {L.}~\bibnamefont
  {Balents}},\ }\bibfield  {title} {\bibinfo {title} {Spin liquids in
  frustrated magnets},\ }\href {https://doi.org/10.1038/nature08917} {\bibfield
   {journal} {\bibinfo  {journal} {Nature}\ }\textbf {\bibinfo {volume}
  {464}},\ \bibinfo {pages} {199} (\bibinfo {year} {2010})}\BibitemShut
  {NoStop}%
\bibitem [{\citenamefont {Savary}\ and\ \citenamefont
  {Balents}(2016{\natexlab{b}})}]{7}%
  \BibitemOpen
  \bibfield  {author} {\bibinfo {author} {\bibfnamefont {L.}~\bibnamefont
  {Savary}}\ and\ \bibinfo {author} {\bibfnamefont {L.}~\bibnamefont
  {Balents}},\ }\bibfield  {title} {\bibinfo {title} {Quantum spin liquids: a
  review},\ }\href {https://doi.org/10.1088/0034-4885/80/1/016502} {\bibfield
  {journal} {\bibinfo  {journal} {Reports on Progress in Physics}\ }\textbf
  {\bibinfo {volume} {80}},\ \bibinfo {pages} {016502} (\bibinfo {year}
  {2016}{\natexlab{b}})}\BibitemShut {NoStop}%
\bibitem [{\citenamefont {Pollmann}\ \emph {et~al.}(2010)\citenamefont
  {Pollmann}, \citenamefont {Turner}, \citenamefont {Berg},\ and\ \citenamefont
  {Oshikawa}}]{8}%
  \BibitemOpen
  \bibfield  {author} {\bibinfo {author} {\bibfnamefont {F.}~\bibnamefont
  {Pollmann}}, \bibinfo {author} {\bibfnamefont {A.~M.}\ \bibnamefont
  {Turner}}, \bibinfo {author} {\bibfnamefont {E.}~\bibnamefont {Berg}},\ and\
  \bibinfo {author} {\bibfnamefont {M.}~\bibnamefont {Oshikawa}},\ }\bibfield
  {title} {\bibinfo {title} {Entanglement spectrum of a topological phase in
  one dimension},\ }\href {https://doi.org/10.1103/PhysRevB.81.064439}
  {\bibfield  {journal} {\bibinfo  {journal} {Phys. Rev. B}\ }\textbf {\bibinfo
  {volume} {81}},\ \bibinfo {pages} {064439} (\bibinfo {year}
  {2010})}\BibitemShut {NoStop}%
\bibitem [{\citenamefont {Jiang}\ \emph {et~al.}(2012)\citenamefont {Jiang},
  \citenamefont {Wang},\ and\ \citenamefont {Balents}}]{9}%
  \BibitemOpen
  \bibfield  {author} {\bibinfo {author} {\bibfnamefont {H.-C.}\ \bibnamefont
  {Jiang}}, \bibinfo {author} {\bibfnamefont {Z.}~\bibnamefont {Wang}},\ and\
  \bibinfo {author} {\bibfnamefont {L.}~\bibnamefont {Balents}},\ }\bibfield
  {title} {\bibinfo {title} {Identifying topological order by entanglement
  entropy},\ }\href {https://doi.org/10.1038/nphys2465} {\bibfield  {journal}
  {\bibinfo  {journal} {Nature Physics}\ }\textbf {\bibinfo {volume} {8}},\
  \bibinfo {pages} {902} (\bibinfo {year} {2012})}\BibitemShut {NoStop}%
\bibitem [{\citenamefont {Haug}\ \emph {et~al.}(2020)\citenamefont {Haug},
  \citenamefont {Amico}, \citenamefont {Kwek}, \citenamefont {Munro},\ and\
  \citenamefont {Bastidas}}]{9b}%
  \BibitemOpen
  \bibfield  {author} {\bibinfo {author} {\bibfnamefont {T.}~\bibnamefont
  {Haug}}, \bibinfo {author} {\bibfnamefont {L.}~\bibnamefont {Amico}},
  \bibinfo {author} {\bibfnamefont {L.-C.}\ \bibnamefont {Kwek}}, \bibinfo
  {author} {\bibfnamefont {W.~J.}\ \bibnamefont {Munro}},\ and\ \bibinfo
  {author} {\bibfnamefont {V.~M.}\ \bibnamefont {Bastidas}},\ }\bibfield
  {title} {\bibinfo {title} {Topological pumping of quantum correlations},\
  }\href {https://doi.org/10.1103/PhysRevResearch.2.013135} {\bibfield
  {journal} {\bibinfo  {journal} {Phys. Rev. Research}\ }\textbf {\bibinfo
  {volume} {2}},\ \bibinfo {pages} {013135} (\bibinfo {year}
  {2020})}\BibitemShut {NoStop}%
\bibitem [{\citenamefont {Bardarson}\ \emph {et~al.}(2012)\citenamefont
  {Bardarson}, \citenamefont {Pollmann},\ and\ \citenamefont {Moore}}]{10}%
  \BibitemOpen
  \bibfield  {author} {\bibinfo {author} {\bibfnamefont {J.~H.}\ \bibnamefont
  {Bardarson}}, \bibinfo {author} {\bibfnamefont {F.}~\bibnamefont
  {Pollmann}},\ and\ \bibinfo {author} {\bibfnamefont {J.~E.}\ \bibnamefont
  {Moore}},\ }\bibfield  {title} {\bibinfo {title} {Unbounded growth of
  entanglement in models of many-body localization},\ }\href
  {https://doi.org/10.1103/PhysRevLett.109.017202} {\bibfield  {journal}
  {\bibinfo  {journal} {Phys. Rev. Lett.}\ }\textbf {\bibinfo {volume} {109}},\
  \bibinfo {pages} {017202} (\bibinfo {year} {2012})}\BibitemShut {NoStop}%
\bibitem [{\citenamefont {Jurcevic}\ \emph {et~al.}(2014)\citenamefont
  {Jurcevic}, \citenamefont {Lanyon}, \citenamefont {Hauke}, \citenamefont
  {Hempel}, \citenamefont {Zoller}, \citenamefont {Blatt},\ and\ \citenamefont
  {Roos}}]{13}%
  \BibitemOpen
  \bibfield  {author} {\bibinfo {author} {\bibfnamefont {P.}~\bibnamefont
  {Jurcevic}}, \bibinfo {author} {\bibfnamefont {B.~P.}\ \bibnamefont
  {Lanyon}}, \bibinfo {author} {\bibfnamefont {P.}~\bibnamefont {Hauke}},
  \bibinfo {author} {\bibfnamefont {C.}~\bibnamefont {Hempel}}, \bibinfo
  {author} {\bibfnamefont {P.}~\bibnamefont {Zoller}}, \bibinfo {author}
  {\bibfnamefont {R.}~\bibnamefont {Blatt}},\ and\ \bibinfo {author}
  {\bibfnamefont {C.~F.}\ \bibnamefont {Roos}},\ }\bibfield  {title} {\bibinfo
  {title} {Quasiparticle engineering and entanglement propagation in a quantum
  many-body system},\ }\href {https://doi.org/10.1038/nature13461} {\bibfield
  {journal} {\bibinfo  {journal} {Nature}\ }\textbf {\bibinfo {volume} {511}},\
  \bibinfo {pages} {202} (\bibinfo {year} {2014})}\BibitemShut {NoStop}%
\bibitem [{\citenamefont {Friis}\ \emph {et~al.}(2018)\citenamefont {Friis},
  \citenamefont {Marty}, \citenamefont {Maier}, \citenamefont {Hempel},
  \citenamefont {Holz\"apfel}, \citenamefont {Jurcevic}, \citenamefont
  {Plenio}, \citenamefont {Huber}, \citenamefont {Roos}, \citenamefont
  {Blatt},\ and\ \citenamefont {Lanyon}}]{14}%
  \BibitemOpen
  \bibfield  {author} {\bibinfo {author} {\bibfnamefont {N.}~\bibnamefont
  {Friis}}, \bibinfo {author} {\bibfnamefont {O.}~\bibnamefont {Marty}},
  \bibinfo {author} {\bibfnamefont {C.}~\bibnamefont {Maier}}, \bibinfo
  {author} {\bibfnamefont {C.}~\bibnamefont {Hempel}}, \bibinfo {author}
  {\bibfnamefont {M.}~\bibnamefont {Holz\"apfel}}, \bibinfo {author}
  {\bibfnamefont {P.}~\bibnamefont {Jurcevic}}, \bibinfo {author}
  {\bibfnamefont {M.~B.}\ \bibnamefont {Plenio}}, \bibinfo {author}
  {\bibfnamefont {M.}~\bibnamefont {Huber}}, \bibinfo {author} {\bibfnamefont
  {C.}~\bibnamefont {Roos}}, \bibinfo {author} {\bibfnamefont {R.}~\bibnamefont
  {Blatt}},\ and\ \bibinfo {author} {\bibfnamefont {B.}~\bibnamefont
  {Lanyon}},\ }\bibfield  {title} {\bibinfo {title} {Observation of entangled
  states of a fully controlled 20-qubit system},\ }\href
  {https://doi.org/10.1103/PhysRevX.8.021012} {\bibfield  {journal} {\bibinfo
  {journal} {Phys. Rev. X}\ }\textbf {\bibinfo {volume} {8}},\ \bibinfo {pages}
  {021012} (\bibinfo {year} {2018})}\BibitemShut {NoStop}%
\bibitem [{\citenamefont {Daley}\ \emph {et~al.}(2012)\citenamefont {Daley},
  \citenamefont {Pichler}, \citenamefont {Schachenmayer},\ and\ \citenamefont
  {Zoller}}]{15}%
  \BibitemOpen
  \bibfield  {author} {\bibinfo {author} {\bibfnamefont {A.~J.}\ \bibnamefont
  {Daley}}, \bibinfo {author} {\bibfnamefont {H.}~\bibnamefont {Pichler}},
  \bibinfo {author} {\bibfnamefont {J.}~\bibnamefont {Schachenmayer}},\ and\
  \bibinfo {author} {\bibfnamefont {P.}~\bibnamefont {Zoller}},\ }\bibfield
  {title} {\bibinfo {title} {Measuring entanglement growth in quench dynamics
  of bosons in an optical lattice},\ }\href
  {https://doi.org/10.1103/PhysRevLett.109.020505} {\bibfield  {journal}
  {\bibinfo  {journal} {Phys. Rev. Lett.}\ }\textbf {\bibinfo {volume} {109}},\
  \bibinfo {pages} {020505} (\bibinfo {year} {2012})}\BibitemShut {NoStop}%
\bibitem [{\citenamefont {Laurell}\ \emph {et~al.}(2021)\citenamefont
  {Laurell}, \citenamefont {Scheie}, \citenamefont {Mukherjee}, \citenamefont
  {Koza}, \citenamefont {Enderle}, \citenamefont {Tylczynski}, \citenamefont
  {Okamoto}, \citenamefont {Coldea}, \citenamefont {Tennant},\ and\
  \citenamefont {Alvarez}}]{16}%
  \BibitemOpen
  \bibfield  {author} {\bibinfo {author} {\bibfnamefont {P.}~\bibnamefont
  {Laurell}}, \bibinfo {author} {\bibfnamefont {A.}~\bibnamefont {Scheie}},
  \bibinfo {author} {\bibfnamefont {C.~J.}\ \bibnamefont {Mukherjee}}, \bibinfo
  {author} {\bibfnamefont {M.~M.}\ \bibnamefont {Koza}}, \bibinfo {author}
  {\bibfnamefont {M.}~\bibnamefont {Enderle}}, \bibinfo {author} {\bibfnamefont
  {Z.}~\bibnamefont {Tylczynski}}, \bibinfo {author} {\bibfnamefont
  {S.}~\bibnamefont {Okamoto}}, \bibinfo {author} {\bibfnamefont
  {R.}~\bibnamefont {Coldea}}, \bibinfo {author} {\bibfnamefont {D.~A.}\
  \bibnamefont {Tennant}},\ and\ \bibinfo {author} {\bibfnamefont
  {G.}~\bibnamefont {Alvarez}},\ }\bibfield  {title} {\bibinfo {title}
  {Quantifying and controlling entanglement in the quantum magnet
  ${\mathrm{cs}}_{2}{\mathrm{cocl}}_{4}$},\ }\href
  {https://doi.org/10.1103/PhysRevLett.127.037201} {\bibfield  {journal}
  {\bibinfo  {journal} {Phys. Rev. Lett.}\ }\textbf {\bibinfo {volume} {127}},\
  \bibinfo {pages} {037201} (\bibinfo {year} {2021})}\BibitemShut {NoStop}%
\bibitem [{\citenamefont {Wootters}(1998)}]{E4-1}%
  \BibitemOpen
  \bibfield  {author} {\bibinfo {author} {\bibfnamefont {W.~K.}\ \bibnamefont
  {Wootters}},\ }\bibfield  {title} {\bibinfo {title} {Entanglement of
  formation of an arbitrary state of two qubits},\ }\href
  {https://doi.org/10.1103/PhysRevLett.80.2245} {\bibfield  {journal} {\bibinfo
   {journal} {Phys. Rev. Lett.}\ }\textbf {\bibinfo {volume} {80}},\ \bibinfo
  {pages} {2245} (\bibinfo {year} {1998})}\BibitemShut {NoStop}%
\bibitem [{\citenamefont {Ollivier}\ and\ \citenamefont {Zurek}(2001)}]{52}%
  \BibitemOpen
  \bibfield  {author} {\bibinfo {author} {\bibfnamefont {H.}~\bibnamefont
  {Ollivier}}\ and\ \bibinfo {author} {\bibfnamefont {W.~H.}\ \bibnamefont
  {Zurek}},\ }\bibfield  {title} {\bibinfo {title} {Quantum discord: A measure
  of the quantumness of correlations},\ }\href
  {https://doi.org/10.1103/PhysRevLett.88.017901} {\bibfield  {journal}
  {\bibinfo  {journal} {Phys. Rev. Lett.}\ }\textbf {\bibinfo {volume} {88}},\
  \bibinfo {pages} {017901} (\bibinfo {year} {2001})}\BibitemShut {NoStop}%
\bibitem [{\citenamefont {Sarandy}(2009)}]{53}%
  \BibitemOpen
  \bibfield  {author} {\bibinfo {author} {\bibfnamefont {M.~S.}\ \bibnamefont
  {Sarandy}},\ }\bibfield  {title} {\bibinfo {title} {Classical correlation and
  quantum discord in critical systems},\ }\href
  {https://doi.org/10.1103/PhysRevA.80.022108} {\bibfield  {journal} {\bibinfo
  {journal} {Phys. Rev. A}\ }\textbf {\bibinfo {volume} {80}},\ \bibinfo
  {pages} {022108} (\bibinfo {year} {2009})}\BibitemShut {NoStop}%
\bibitem [{\citenamefont {Kitaev}\ and\ \citenamefont {Preskill}(2006)}]{EE1}%
  \BibitemOpen
  \bibfield  {author} {\bibinfo {author} {\bibfnamefont {A.}~\bibnamefont
  {Kitaev}}\ and\ \bibinfo {author} {\bibfnamefont {J.}~\bibnamefont
  {Preskill}},\ }\bibfield  {title} {\bibinfo {title} {Topological entanglement
  entropy},\ }\href {https://doi.org/10.1103/PhysRevLett.96.110404} {\bibfield
  {journal} {\bibinfo  {journal} {Phys. Rev. Lett.}\ }\textbf {\bibinfo
  {volume} {96}},\ \bibinfo {pages} {110404} (\bibinfo {year}
  {2006})}\BibitemShut {NoStop}%
\bibitem [{\citenamefont {Levin}\ and\ \citenamefont {Wen}(2006)}]{EE2}%
  \BibitemOpen
  \bibfield  {author} {\bibinfo {author} {\bibfnamefont {M.}~\bibnamefont
  {Levin}}\ and\ \bibinfo {author} {\bibfnamefont {X.-G.}\ \bibnamefont
  {Wen}},\ }\bibfield  {title} {\bibinfo {title} {Detecting topological order
  in a ground state wave function},\ }\href
  {https://doi.org/10.1103/PhysRevLett.96.110405} {\bibfield  {journal}
  {\bibinfo  {journal} {Phys. Rev. Lett.}\ }\textbf {\bibinfo {volume} {96}},\
  \bibinfo {pages} {110405} (\bibinfo {year} {2006})}\BibitemShut {NoStop}%
\bibitem [{\citenamefont {Li}\ and\ \citenamefont {Haldane}(2008)}]{EE3}%
  \BibitemOpen
  \bibfield  {author} {\bibinfo {author} {\bibfnamefont {H.}~\bibnamefont
  {Li}}\ and\ \bibinfo {author} {\bibfnamefont {F.~D.~M.}\ \bibnamefont
  {Haldane}},\ }\bibfield  {title} {\bibinfo {title} {Entanglement spectrum as
  a generalization of entanglement entropy: Identification of topological order
  in non-abelian fractional quantum hall effect states},\ }\href
  {https://doi.org/10.1103/PhysRevLett.101.010504} {\bibfield  {journal}
  {\bibinfo  {journal} {Phys. Rev. Lett.}\ }\textbf {\bibinfo {volume} {101}},\
  \bibinfo {pages} {010504} (\bibinfo {year} {2008})}\BibitemShut {NoStop}%
\bibitem [{\citenamefont {Laflorencie}(2016)}]{EE4}%
  \BibitemOpen
  \bibfield  {author} {\bibinfo {author} {\bibfnamefont {N.}~\bibnamefont
  {Laflorencie}},\ }\bibfield  {title} {\bibinfo {title} {Quantum entanglement
  in condensed matter systems},\ }\href
  {https://doi.org/https://doi.org/10.1016/j.physrep.2016.06.008} {\bibfield
  {journal} {\bibinfo  {journal} {Physics Reports}\ }\textbf {\bibinfo {volume}
  {646}},\ \bibinfo {pages} {1} (\bibinfo {year} {2016})},\ \bibinfo {note}
  {quantum entanglement in condensed matter systems}\BibitemShut {NoStop}%
\bibitem [{\citenamefont {Bergschneider}\ \emph {et~al.}(2019)\citenamefont
  {Bergschneider}, \citenamefont {Klinkhamer}, \citenamefont {Becher},
  \citenamefont {Klemt}, \citenamefont {Palm}, \citenamefont {Z\"urn},
  \citenamefont {Jochim},\ and\ \citenamefont {Preiss}}]{EE5}%
  \BibitemOpen
  \bibfield  {author} {\bibinfo {author} {\bibfnamefont {A.}~\bibnamefont
  {Bergschneider}}, \bibinfo {author} {\bibfnamefont {V.~M.}\ \bibnamefont
  {Klinkhamer}}, \bibinfo {author} {\bibfnamefont {J.~H.}\ \bibnamefont
  {Becher}}, \bibinfo {author} {\bibfnamefont {R.}~\bibnamefont {Klemt}},
  \bibinfo {author} {\bibfnamefont {L.}~\bibnamefont {Palm}}, \bibinfo {author}
  {\bibfnamefont {G.}~\bibnamefont {Z\"urn}}, \bibinfo {author} {\bibfnamefont
  {S.}~\bibnamefont {Jochim}},\ and\ \bibinfo {author} {\bibfnamefont {P.~M.}\
  \bibnamefont {Preiss}},\ }\bibfield  {title} {\bibinfo {title} {Experimental
  characterization of two-particle entanglement through position and momentum
  correlations},\ }\href {https://doi.org/10.1038/s41567-019-0508-6} {\bibfield
   {journal} {\bibinfo  {journal} {Nature Physics}\ }\textbf {\bibinfo {volume}
  {15}},\ \bibinfo {pages} {640} (\bibinfo {year} {2019})}\BibitemShut
  {NoStop}%
\bibitem [{\citenamefont {Lanczos}(1950)}]{Lanczos_1950}%
  \BibitemOpen
  \bibfield  {author} {\bibinfo {author} {\bibfnamefont {C.}~\bibnamefont
  {Lanczos}},\ }\bibfield  {title} {\bibinfo {title} {An iteration method for
  the solution of the eigenvalue problem of linear differential and integral
  operators},\ }\href {https://doi.org/10.6028/jres.045.026} {\bibfield
  {journal} {\bibinfo  {journal} {Journal of Research of the National Bureau of
  Standards}\ }\textbf {\bibinfo {volume} {45}},\ \bibinfo {pages} {255}
  (\bibinfo {year} {1950})}\BibitemShut {NoStop}%
\bibitem [{\citenamefont {White}(1992)}]{dmrg1992}%
  \BibitemOpen
  \bibfield  {author} {\bibinfo {author} {\bibfnamefont {S.~R.}\ \bibnamefont
  {White}},\ }\bibfield  {title} {\bibinfo {title} {Density matrix formulation
  for quantum renormalization groups},\ }\href
  {https://doi.org/10.1103/PhysRevLett.69.2863} {\bibfield  {journal} {\bibinfo
   {journal} {Phys. Rev. Lett.}\ }\textbf {\bibinfo {volume} {69}},\ \bibinfo
  {pages} {2863} (\bibinfo {year} {1992})}\BibitemShut {NoStop}%
\bibitem [{\citenamefont {Stoudenmire}\ and\ \citenamefont
  {White}(2012)}]{Stoudenmire2012}%
  \BibitemOpen
  \bibfield  {author} {\bibinfo {author} {\bibfnamefont {E.}~\bibnamefont
  {Stoudenmire}}\ and\ \bibinfo {author} {\bibfnamefont {S.~R.}\ \bibnamefont
  {White}},\ }\bibfield  {title} {\bibinfo {title} {Studying two-dimensional
  systems with the density matrix renormalization group},\ }\href
  {https://doi.org/10.1146/annurev-conmatphys-020911-125018} {\bibfield
  {journal} {\bibinfo  {journal} {Annual Review of Condensed Matter Physics}\
  }\textbf {\bibinfo {volume} {3}},\ \bibinfo {pages} {111} (\bibinfo {year}
  {2012})}\BibitemShut {NoStop}%
\bibitem [{\citenamefont {Troyer}\ and\ \citenamefont
  {Wiese}(2005)}]{Troyer2005}%
  \BibitemOpen
  \bibfield  {author} {\bibinfo {author} {\bibfnamefont {M.}~\bibnamefont
  {Troyer}}\ and\ \bibinfo {author} {\bibfnamefont {U.-J.}\ \bibnamefont
  {Wiese}},\ }\bibfield  {title} {\bibinfo {title} {Computational complexity
  and fundamental limitations to fermionic quantum monte carlo simulations},\
  }\href {https://doi.org/10.1103/PhysRevLett.94.170201} {\bibfield  {journal}
  {\bibinfo  {journal} {Phys. Rev. Lett.}\ }\textbf {\bibinfo {volume} {94}},\
  \bibinfo {pages} {170201} (\bibinfo {year} {2005})}\BibitemShut {NoStop}%
\bibitem [{\citenamefont {Misguich}\ and\ \citenamefont
  {Jolicoeur}(2021)}]{Misguich2021}%
  \BibitemOpen
  \bibfield  {author} {\bibinfo {author} {\bibfnamefont {G.}~\bibnamefont
  {Misguich}}\ and\ \bibinfo {author} {\bibfnamefont {T.}~\bibnamefont
  {Jolicoeur}},\ }\bibfield  {title} {\bibinfo {title} {{DMRG} study of {FQHE}
  systems in the open cylinder geometry},\ }\href
  {https://doi.org/10.1088/1742-6596/1740/1/012043} {\bibfield  {journal}
  {\bibinfo  {journal} {Journal of Physics: Conference Series}\ }\textbf
  {\bibinfo {volume} {1740}},\ \bibinfo {pages} {012043} (\bibinfo {year}
  {2021})}\BibitemShut {NoStop}%
\bibitem [{\citenamefont {Scholl}\ \emph {et~al.}(2021)\citenamefont {Scholl},
  \citenamefont {Schuler}, \citenamefont {Williams}, \citenamefont
  {Eberharter}, \citenamefont {Barredo}, \citenamefont {Schymik}, \citenamefont
  {Lienhard}, \citenamefont {Henry}, \citenamefont {Lang}, \citenamefont
  {Lahaye}, \citenamefont {Läuchli},\ and\ \citenamefont
  {Browaeys}}]{Scholl2021}%
  \BibitemOpen
  \bibfield  {author} {\bibinfo {author} {\bibfnamefont {P.}~\bibnamefont
  {Scholl}}, \bibinfo {author} {\bibfnamefont {M.}~\bibnamefont {Schuler}},
  \bibinfo {author} {\bibfnamefont {H.~J.}\ \bibnamefont {Williams}}, \bibinfo
  {author} {\bibfnamefont {A.~A.}\ \bibnamefont {Eberharter}}, \bibinfo
  {author} {\bibfnamefont {D.}~\bibnamefont {Barredo}}, \bibinfo {author}
  {\bibfnamefont {K.-N.}\ \bibnamefont {Schymik}}, \bibinfo {author}
  {\bibfnamefont {V.}~\bibnamefont {Lienhard}}, \bibinfo {author}
  {\bibfnamefont {L.-P.}\ \bibnamefont {Henry}}, \bibinfo {author}
  {\bibfnamefont {T.~C.}\ \bibnamefont {Lang}}, \bibinfo {author}
  {\bibfnamefont {T.}~\bibnamefont {Lahaye}}, \bibinfo {author} {\bibfnamefont
  {A.~M.}\ \bibnamefont {Läuchli}},\ and\ \bibinfo {author} {\bibfnamefont
  {A.}~\bibnamefont {Browaeys}},\ }\bibfield  {title} {\bibinfo {title}
  {Quantum simulation of 2d antiferromagnets with hundreds of rydberg atoms},\
  }\href {https://doi.org/10.1038/s41586-021-03585-1} {\bibfield  {journal}
  {\bibinfo  {journal} {Nature}\ }\textbf {\bibinfo {volume} {595}},\ \bibinfo
  {pages} {233} (\bibinfo {year} {2021})}\BibitemShut {NoStop}%
\bibitem [{\citenamefont {Samajdar}\ \emph {et~al.}(2021)\citenamefont
  {Samajdar}, \citenamefont {Ho}, \citenamefont {Pichler}, \citenamefont
  {Lukin},\ and\ \citenamefont {Sachdev}}]{Samajdar2021}%
  \BibitemOpen
  \bibfield  {author} {\bibinfo {author} {\bibfnamefont {R.}~\bibnamefont
  {Samajdar}}, \bibinfo {author} {\bibfnamefont {W.~W.}\ \bibnamefont {Ho}},
  \bibinfo {author} {\bibfnamefont {H.}~\bibnamefont {Pichler}}, \bibinfo
  {author} {\bibfnamefont {M.~D.}\ \bibnamefont {Lukin}},\ and\ \bibinfo
  {author} {\bibfnamefont {S.}~\bibnamefont {Sachdev}},\ }\bibfield  {title}
  {\bibinfo {title} {Quantum phases of rydberg atoms on a kagome lattice},\
  }\bibfield  {journal} {\bibinfo  {journal} {Proceedings of the National
  Academy of Sciences}\ }\textbf {\bibinfo {volume} {118}},\ \href
  {https://doi.org/10.1073/pnas.2015785118} {10.1073/pnas.2015785118} (\bibinfo
  {year} {2021})\BibitemShut {NoStop}%
\bibitem [{\citenamefont {Fishman}\ \emph {et~al.}(2020)\citenamefont
  {Fishman}, \citenamefont {White},\ and\ \citenamefont
  {Stoudenmire}}]{itensor}%
  \BibitemOpen
  \bibfield  {author} {\bibinfo {author} {\bibfnamefont {M.}~\bibnamefont
  {Fishman}}, \bibinfo {author} {\bibfnamefont {S.~R.}\ \bibnamefont {White}},\
  and\ \bibinfo {author} {\bibfnamefont {E.~M.}\ \bibnamefont {Stoudenmire}},\
  }\href@noop {} {\bibinfo {title} {The \mbox{ITensor} software library for
  tensor network calculations}} (\bibinfo {year} {2020}),\ \Eprint
  {https://arxiv.org/abs/2007.14822} {arXiv:2007.14822} \BibitemShut {NoStop}%
\end{thebibliography}%

\end{document}